\documentclass[a4paper,11pt]{article}
\pdfoutput=1
\usepackage{jheppub}
\usepackage[T1]{fontenc} 

\newcommand{\ph}{t}
\newcommand{\qm}{\ell}
\newcommand{\qmn}{\boldsymbol{n}}
\newcommand{\im}{\mathrm{Im}}
\newcommand{\re}{\mathrm{Re}}
\newcommand{\MeV}{\;\text{MeV}}
\newcommand{\ri}{\mathrm{i}}
\newcommand{\rd}{\mathrm{d}}
\newcommand{\rme}{\mathrm{e}}
\newcommand{\Nc}{N_\text{c}}

\newcommand{\bzero}{\boldsymbol{0}}
\newcommand{\bx}{\boldsymbol{x}}

\newcommand{\bk}{\boldsymbol{k}}
\newcommand{\bp}{\boldsymbol{p}}

\newcommand{\bB}{\boldsymbol{B}}
\newcommand{\Slash}[1]{\ooalign{\hfil/\hfil\crcr$#1$}}
\newcommand{\Dn}{S}
\newcommand{\calC}{\mathcal{C}}
\newcommand{\calE}{\mathcal{E}}
\newcommand{\calM}{\mathcal{M}}
\newcommand{\calL}{\mathcal{L}}
\newcommand{\calP}{\mathcal{P}}
\newcommand{\calS}{\mathcal{S}}
\newcommand{\calQ}{\mathcal{Q}}
\newcommand{\calT}{\mathcal{T}}
\newcommand{\calJ}{\mathcal{J}}

\newcommand{\tr}{\mathrm{tr}\,}

\newcommand{\sgn}{\mathop{\mathrm{sgn}}}
\newcommand{\Pmu}{P}
\newcommand{\Ffunc}{\alpha}
\newcommand{\Gfunc}{\beta}
\newcommand{\feq}{f_{\text{eq}}}
\newcommand{\fbareq}{\bar{f}_{\text{eq}}}
\newcommand{\gequiv}{g_{\text{eq}}}

\title{Resummation for the Field-theoretical Derivation
  of the Negative Magnetoresistance}

\preprint{RIKEN-QHP-412}

\author[a,b]{Kenji Fukushima}
\author[d,e]{Yoshimasa Hidaka}
\affiliation[a]{Department of Physics, The University of Tokyo,\\
  7-3-1 Hongo, Bunkyo-ku, Tokyo 113-0033, Japan}
\affiliation[b]{Institute for Physics of Intelligence, The University of Tokyo,\\
  7-3-1 Hongo, Bunkyo-ku, Tokyo 113-0033, Japan}
\affiliation[d]{Theoretical Research Division, Nishina Center,
                RIKEN,\\ 2-1 Hirosawa, Wako, Saitama 351-0198, Japan}
\affiliation[e]{RIKEN iTHEMS,
                RIKEN,\\ 2-1 Hirosawa, Wako, Saitama 351-0198, Japan}
\emailAdd{fuku@nt.phys.s.u-tokyo.ac.jp}
\emailAdd{hidaka@riken.jp}

\abstract{
  We show detailed derivation of the electric conductivity of quark
  matter at finite temperature and density under a magnetic field.
  We especially focus on the longitudinal electric conductivity along
  the magnetic direction and establish the field-theoretical
  description of the negative magnetoresistance as observed in chiral
  materials.  With increasing magnetic field our microscopic
  calculation leads to changing behavior from approximately quadratic
  to asymptotically linear dependence of the electric conductivity,
  while the magnetic dependence is quadratic in the conventional
  relaxation time approximation.  The presented formulation founds a
  firm basis for the physical interpretation of the negative
  magnetoresistance in terms of the particle and the hydrodynamic
  contributions, as well as it offers general methodology applicable
  for various transport coefficients.
}

\begin{document}
\maketitle
\flushbottom

\section{Introduction}
\label{sec:intro}

Chiral magnetic effect (CME) was originally proposed as a measurable
probe to topologically nontrivial excitations in relativistic
nucleus-nucleus collisions~\cite{Kharzeev:2007jp}.  The theoretical
framework has been elaborated with recognition of ``chiral chemical
potential'' introduced later in Ref.~\cite{Fukushima:2008xe}.  In the
formulation with the chiral chemical potential the theoretical link
between the CME and the chiral anomaly in quantum chromodynamics (QCD)
has become transparent in a similar fashion to an axial counterpart
that had been known~\cite{Metlitski:2005pr} and is called the chiral
separation effect nowadays.   The idea can be translated to simpler
setups with the chiral chemical potential replaced by electromagnetic
background fields~\cite{Fukushima:2010vw,Son:2012bg}, which has opened
a new opportunity to observe the chiral anomaly not necessarily in QCD
but in more accessible experiments~\cite{Li:2014bha}.  For a fairly
complete list of references (especially earlier works), see a recent
review~\cite{Kharzeev:2015znc}.  Readers can also consult an essay of
Ref.~\cite{Fukushima:2012vr} for developments of the idea.

One clear signature for the CME is supposed to be the negative
magnetoresistance as proposed in Ref.~\cite{Son:2012bg} and it has
been confirmed experimentally in Ref.~\cite{Kharzeev:2015znc} followed
by independent experiments with Dirac
semimetals~\cite{Xiong413,Li2015} and Weyl
semimetals~\cite{2015Huang,2016Arnold}.  The negative
magnetoresistance refers to decreasing electric resistance, $\rho$, or
equivalently increasing electric conductivity, $\sigma=1/\rho$, along
the magnetic direction with increasing magnetic field $B$.
Intuitively, as derived in Ref.~\cite{Kharzeev:2015znc}, one can
understand the physical mechanism as follows:  The chirality
production is enhanced as $\propto EB$ by the chiral anomaly through
the Schwinger mechanism in the presence of parallel electromagnetic
fields~\cite{Fukushima:2010vw,Dunne:2004nc,Copinger:2018ftr} (see also
Ref.~\cite{Fukushima:2018grm} for a review), and successively, the
chiral magnetic effect adds an extra conductivity term
$\propto (EB) B \propto B^2$.

Although the above-mentioned mechanism is robust, one may think that
$\rho$ or $\sigma$ could be anyway affected by $B$.  For instance, a
hot-QCD medium can be approximated as a gas of hadron resonances, in
which, superficially, the chiral anomaly is irrelevant.  Nevertheless,
it has been revealed that the electric charge fluctuation in this model is
significantly enhanced with increasing $B$ especially at finite
chemical potential $\mu$~\cite{Fukushima:2016vix}, which also suggests
significant enhancement in the electric conductivity.  One might be then
tempted to make some unfavorable statements about the
magnetoresistance and the chiral anomaly.  In principle, however, the
model could implicitly incorporate the effect of the chiral anomaly,
and it is impossibly difficult to draw a solid conclusion from such
model studies and crude approximations.

To establish the physical interpretation, therefore, it is desirable
to complete field-theoretical calculations for $\rho$ or $\sigma$
taking account of full nonlinear dependence on $B$.  The starting
point for the conductivity calculation should be the Kubo formula, but
as we will discuss in great details, some resummation is quite often
indispensable and literal calculations might be formidably
cumbersome.  Then, it would be one of the most tractable approaches to
utilize the kinetic equation, as adopted also in
Ref.~\cite{Son:2012bg}.  In the estimate previously made in
Ref.~\cite{Son:2012bg}, however, the Chiral
Kinetic Theory (CKT)~\cite{Son:2012wh,Stephanov:2012ki,Gao:2012ix,
Son:2012zy,Chen:2012ca,Manuel:2013zaa,Chen:2014cla,Chen:2015gta,
Hidaka:2016yjf,Hidaka:2017auj,Hidaka:2018ekt,Mueller:2017arw,
Mueller:2017lzw,Huang:2018wdl,Carignano:2018gqt,Dayi:2018xdy,
Liu:2018xip,Lin:2019ytz} was assumed, in
which the Dirac spin structure reduces to the Berry curvature under the
adiabatic approximation.  Besides, in Ref.~\cite{Son:2012bg} and in
the theoretical argument in Ref.~\cite{Kharzeev:2015znc}, the
relaxation-time approximation was used with a constant relaxation
time.  These treatments are appropriate for small $B$, but are not
necessarily so for large $B$.

If $B$ (multiplied by a typical electric charge) is the largest energy
scale in systems with charged particles, the lowest Landau level (LLL)
would make a dominant contribution to physical observables.  The
theoretical calculation is then feasible enough in the approximation to take
the LLL only (i.e., LLLA).  Moreover, we point out that the whole
theoretical description would be far cleaner in QCD matter rather than
in condensed matter systems;  for the former the fundamental theory is
QCD, but for the the latter quantum electrodynamics (QED) cannot be
directly employed.  In fact, various transport coefficients in QCD
matter have been estimated in the LLLA of QCD that is justified with
assumed hierarchy, $\sqrt{qB} \gg T \gg gT$, where $q$ is an electric
charge and $g$ is a QCD charge of quarks.  The heavy-quark diffusion
constant is found in Ref.~\cite{Fukushima:2015wck}, the electric
conductivity in Refs.~\cite{Hattori:2016cnt,Hattori:2016lqx}, and the
viscosity in Ref.~\cite{Hattori:2017qih}, etc.

Usually the LLLA is a good approximation as long as the magnetic scale
is large enough.  For the longitudinal electric conductivity, however,
the LLLA is broken down due to phase space and it is insufficient for
the negative magnetoresistance at intermediate strength of the
magnetic field.  As pointed out in Ref.~\cite{Fukushima:2015wck},
strong-$B$ fields reduce quark dispersions from (3+1)- to
(1+1)-dimensional dynamics, and quarks thus cannot scatter if they are
massless; otherwise, the energy-momentum conservations would be
violated.  This implies that the electric conductivity in the LLLA
diverges as $m_f$, a mass of $f$-flavor quark, goes to zero, i.e.,
$\sigma_{\rm LLLA}(m_{f}\to 0)\to\infty$.  We will
circumvent diverging $\sigma$ by considering higher Landau levels.
Specifically, in the present work, we assume a different regime of
relevant scales, i.e., $T \gtrsim \sqrt{qB} \gg gT$, which is more
relevant for realistic magnetic situations.  In this regime the
magnetic field is no longer the largest scale, but still we require
the last inequality in order to neglect thermal excitation effects as
compared to the magnetically induced gap.
Here let us make a comment on a subtle point about the massless
limit.  As we will closely discuss later, in the strict massless
limit, the axial charge increases linearly with time, and the physical
conductivity diverges even beyond the LLLA{}.  Nevertheless, the
transport coefficient used for the hydrodynamics (in which the
conserved axial charge makes a hydrodynamic mode) is still finite,
that is what we are computing in this work.  The point is the time
scale;  if we take the long time (i.e., small frequency) limit first
and approach the massless limit, the conductivity would diverge.  If
we first adopt small masses and then take the hydrodynamic limit, the
conductivity would remain finite as we will demonstrate.  Such a
seeming difference appears from two hydrodynamic regimes, and the
hydrodynamical evolution in respective frameworks should lead to the
same renormalized conductivity.

We would note one more advantage of performing the calculation in QCD
on top of its being a theoretically clean setup.  There are some
numerical results available from first-principles simulations by means
of lattice discretized formulation of QCD, namely, the lattice-QCD
simulations.    The latest quenched lattice-QCD results are
$0.2 C_{\rm em} \lesssim \sigma/T \lesssim 0.7 C_{\rm em}$ for
$T=1.1\,T_{\rm c}$ where $C_{\rm em} :=\sum_f q_f^2$ and
$T_{\rm c}$ is the QCD critical temperature~\cite{Ding:2016hua}.  See
also Refs.~\cite{Gupta:2003zh,Aarts:2007wj,Ding:2010ga} for earlier
results.  These estimated values are from the simulations at $T\neq 0$
and $B=0$, but the order of magnitude is expected to be the same as
our calculations for $T \gtrsim \sqrt{qB} \gg gT$ if the magnetic
dependence of $\sigma$ is smooth enough.  The consistency would
support validity of our calculations and we will numerically confirm
this expectation later.

Besides physics interest, as we will discuss in this work, the
theoretical framework of the resummation scheme is very interesting on
its own.  There are two major
technical advances reported in this paper.  One is a clear
prescription for the hydrodynamic zero modes that appear at finite
density associated with conserved quantities.  The other is a
field-theoretical derivation of the Boltzmann equations as a result of
the resummation on the vertex.  Since the latter particularly needs
judicious transformations of expressions, we first discuss physics
parts based on the assumed Boltzmann equations, and then proceed into
their step-by-step derivations.

This paper is organized as follows.  In Sec.~\ref{sec:definitions} we
summarize our conventions and field-theoretical definitions of
physical quantities of our interest.  As warm-up exercises in
Sec.~\ref{sec:transverse} we perform explicit calculations for the
Hall conductivity and give discussions on the transverse conductivity
which is vanishing at the one-loop order.
Section~\ref{sec:longitudinal} is the central part of this paper
consisting of subsections devoted to the methods, the numerical
results, and the physics discussions.  To fill in the gap between the
Kubo formula and the Boltzmann equations, we explicate step-by-step
derivations in Sec.~\ref{sec:diagram}.  We finally make conclusions in
Sec.~\ref{sec:summary}.

\section{Conventions, definitions, and the final results first}
\label{sec:definitions}

For readers to have better accessibility to the contents, we will show
our final results first here.  Before doing this, let us summarize the
conventions and the definitions used in this work.  Bearing them in
mind, readers can understand which physical quantity we are
calculating and what the final results are, before the subsequent
technical discussions.  We will gradually proceed to technical details
in later sections, first the one-loop integral in
Sec.~\ref{sec:transverse}, next the resummed kinetic description in
Sec.~\ref{sec:longitudinal}, and finally the full diagrammatic
derivation in Sec.~\ref{sec:diagram}.

Our convention of the Lorentz four vector assumes the Minkowski
metric, $\eta_{\mu\nu}=\mathrm{diag}(1,-1,-1,-1)$.  We choose the
covariant derivative as $D_\mu=\partial_\mu+\ri q_f A_\mu$ where $q_f$
is the electric charge carried by quark flavor $f$ and $A_\mu$ is the
electromagnetic gauge field.  In this paper, we treat the gluonic
degrees of freedom as well, but the background field is only
electromagnetic.  We often use a sloppy but intuitive notation for the
contravariant components of the four momentum vector as
$p^\mu=(p^0, p^1, p^2, p^3)=(\varepsilon, p_x, p_y, p_z)$ using the
coordinates $x$, $y$, and $z$, and sometimes switch the latter back to
the former for notational convenience.

The physical observable of our utmost interest in this work is the
electric conductivity, which is given by the Kubo formula as follows:
\begin{equation}
  \sigma^{ij}  = \lim_{k_0\to0} \lim_{\bk\to\bzero}
  \frac{1}{2\ri k_0} \bigl[ \Pi_R^{ij}(k) - \Pi_A^{ij}(k) \bigr]
  \label{eq:Kubo}
\end{equation}
with $i$, $j$ taking spatial coordinates, $x$, $y$, $z$.  Here, the
order of two limits is important;  $\bk\to \bzero$ is taken first and
$k_0\to 0$ next.  In the above expression $\Pi_R^{ij}(k)$ and
$\Pi_A^{ij}(k)$ are the retarded and advanced polarization functions,
respectively, defined by
\begin{align}
  \Pi_R^{ij}(k) &:= \ri \int \rd^4 x\, \rme^{\ri k\cdot x}\,
  \theta(t) \bigl\langle [j^i (x),j^j (0)] \bigr\rangle\,,\\
  \Pi_A^{ij}(k) &:= -\ri \int \rd^4 x\, \rme^{\ri k\cdot x}\,
  \theta(-t) \bigl\langle [j^i (x),j^j (0)] \bigr\rangle\,.
\end{align}
It is crucially important to note that $j^i$ is not exactly the
electric current if a finite density is coupled.  This is so because,
as we will see later, zero modes or hydrodynamic modes make the
inversion operation ill-defined and we should subtract them (which
should be included in dynamics of hydrodynamic modes with the
calculated conductivity, not in the conductivity itself).
Specifically, the above $j^i$ should be given by
\begin{equation}
  j^i = j_{\rm em}^i - \frac{n_e T^{0i}}{\calE+\calP_i} \,,
\label{eq:jsubt}
\end{equation}
where $T^{\mu\nu}$ represents the energy momentum tensor, $n_e$ the
net electric charge density, $\calE:=\langle T^{00}\rangle$ the energy
density, and $\calP_i:=\langle T^{ii}\rangle$ the pressure.  Because
$j^i$ is an operator, the numerator of the correction term is an
operator $\propto T^{0i}$, while the denominator is an expectation value,
namely, the enthalpy of the system.  In QCD with multiple quark flavors
the electric current is a sum of contributions from all flavors $f$, i.e.,
\begin{equation}
  j_{\rm em}^i = \sum_f q_f \bar{\psi}_f \gamma^i \psi_f \,.
\end{equation}
If the net electric charge is zero (as is the case at zero density),
$j^i=j_{\rm em}^i$ simply holds.

Now, let us explain the decomposition of the electric conductivity
tensor into the longitudinal component and others.  The external
magnetic field introduces a preferred direction, and we decompose the
anisotropic tensor structure using $\hat{B}^i:=B^i/|\bB|$ as
\begin{equation}
  \sigma^{ij} = \sigma_H\,\epsilon^{ijk}\hat{B}^k
  + \sigma_\parallel\,\hat{B}^i \hat{B}^j
  + \sigma_\perp\,(\delta^{ij}-\hat{B}^i \hat{B}^j)\,,
\end{equation}
where $\sigma_H$, $\sigma_\parallel$, and $\sigma_\perp$ represent the
Hall conductivity, the longitudinal conductivity, and the transverse
conductivity, respectively.  Without loss of generality we can
identify the magnetic field direction with the $z$ axis, so that we
explicitly identify these conductivities as
$\sigma_H=\sigma^{12}=-\sigma^{21}$, $\sigma_\parallel=\sigma^{33}$,
and $\sigma_\perp=\sigma^{11}=\sigma^{22}$.

So far, all the formulas are general, and we do not impose any
physical condition yet.  In this paper, as we declared in the
introduction, we work only in the weak coupling case and mostly
consider a regime in which the following hierarchy is satisfied:
\begin{equation}
  T \;\gtrsim\; \sqrt{|q_f B|} \;\gg\; gT \,,
  \label{eq:hierarchy}
\end{equation}
where $f$ is an arbitrary flavor, $g$ is the quark-gluon coupling
constant (which is assumed to be small in the weak coupling case), and
$T$ is the temperature.  We note that the above hierarchy is not a
sufficient condition for the LLLA because of the phase space
suppression for light fermions, as was also emphasized in the
introduction.  It is, therefore, important to take account of full
Landau levels as we will do in this work.  One may then wonder why
such a lower bound in Eq.~\eqref{eq:hierarchy} is needed though we
consider all the Landau levels.  This is because, as we will
explicitly see later, we make an approximation not to include the
finite-$T$ induced self-energy $\sim g^2 T^2$ in the quark propagator,
which is justified if the $B$ induced gap is much larger, i.e.,
$|q_f B|\gg g^2 T^2$ that is nothing but the second inequality in
Eq.~\eqref{eq:hierarchy}.  A side remark is that, as argued in
Ref.~\cite{Fukushima:2015wck}, the lowest Landau level is not gapped
with $|q_f B|$, but still, no finite-$T$ term appears at all in (1+1)
dimensions, so that our approximated quark propagator can be valid.
In contrast, the upper bound by $T$ has a looser origin.  Later, we
will limit our calculation to the lowest order perturbation in $g$
(i.e., $1\leftrightarrow 2$ processes), but if $B$ becomes too larger
than $T$, higher order processes might not be dropped.  To stay in the
safe side we impose this upper bound.  Once the scattering processes
are fixed, however, the calculation itself does not break down even
for arbitrarily large $B$, and this large-$B$ analysis is what we will
perform in our discussions on the LLLA later. 

We make a short remark about the sphaleron transition which may in
principle contribute to the electric conductivity.  In QCD at weak
coupling, however, the sphaleron transition rate is as small as
$\alpha_s^5$, and we can safely neglect it.  Thus, in this work, we
focus on the perturbative contribution only.

Now that we articulated what we want to calculate under which
condition, then, we shall present our final results in advance.   The
easiest to obtain is the Hall conductivity, that is,
\begin{equation}
  \sigma_H = \frac{n_e}{B}\,,
  \label{eq:Hall}
\end{equation}
as will be elaborated in the next warm-up section, where we will deal
with explicit expressions for the propagators in the magnetic field.
The transverse conductivity is vanishing at the one-loop level and, as
discussed in the next section, higher-loop terms are suppressed by the
magnetic field parametrically as
\begin{equation}
  \frac{\sigma_\perp}{T} \;\sim\; \frac{g^2 T^2}{|q_f B|}\,.
 \label{eq:sigmaperp}
\end{equation}
Since the transverse conductivity is small in our hierarchy
regime~\eqref{eq:hierarchy}, we would not try to quantify
$\sigma_\perp$ any further.

\begin{figure}
  \centering 
  \includegraphics[width=0.7\textwidth]{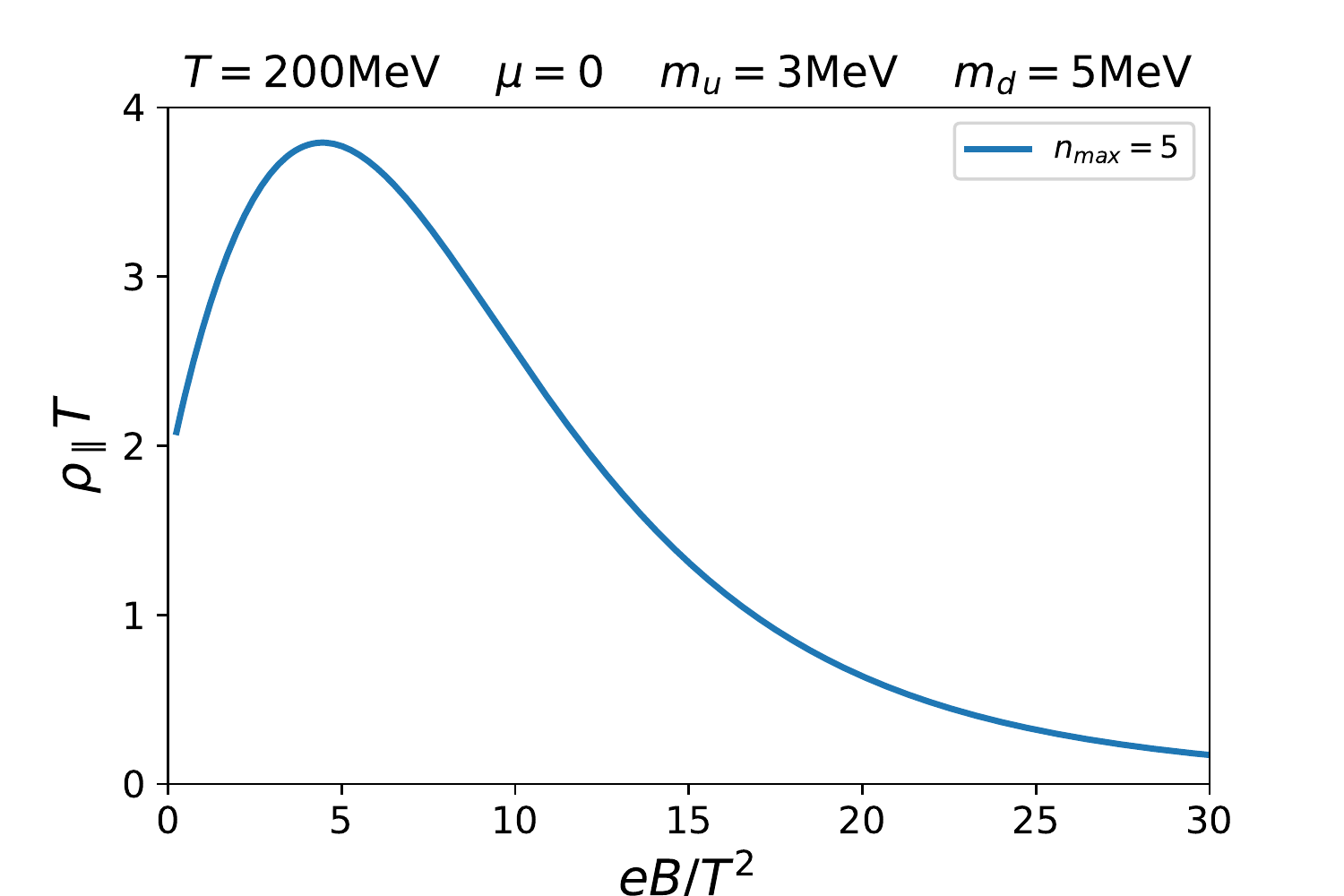}
  \caption{Negative magnetoresistance;  $\rho_\parallel$ decreases
    with increasing magnetic field $eB$.  The temperature is
    $T=200\MeV$ and the density is zero, and $n_{\rm max}=5$
    represents the highest Landau level taken into the calculation.}
  \label{fig:BDepLarge_rho}
\end{figure}

The most interesting is the longitudinal electric conductivity
$\sigma_\parallel$ or the resistance $\rho_\parallel$, which is made
dimensionless with the temperature as
\begin{equation}
  \rho_\parallel T := \frac{1}{(\sigma_\parallel / T)}\,. 
\end{equation}
The negative magnetoresistance signifies decreasing behavior of
$\rho_\parallel$ with increasing magnetic field, and this is precisely
what we finally find in our calculation as shown in
Fig.~\ref{fig:BDepLarge_rho}.  We note that $e$ of $eB$ on the
horizontal axis is the (positive) elementary charge, and in this
calculation, $u$ and $d$ quark flavors are included
with $q_u=\frac{2}{3}e$ and $q_d=-\frac{1}{3}e$.
As is clear from the plot, we have
confirmed the negative magnetoresistance from field-theoretical
calculations without assuming anything special about the chirality
production and the topological transport.  Here, we should emphasize
that no assumption does not mean no chiral anomaly.  Actually, if one
solves the Dirac equation in the presence of electromagnetic
background, the chiral anomaly should be automatically incorporated.
In gauge theories fermions are bilinear, so that fermion integrals can
be one-loop exact apart from gauge fluctuations.

\section{Warm-ups : the Hall conductivity and the transverse conductivity}
\label{sec:transverse}

For diagrammatic calculations the elementary building blocks are the
propagator and the vertices.  In the presence of the magnetic field
even the free propagator takes a cumbersome form, as we cope with in
this section.  We will discuss the resummed vertices in the next
section.
Here, we will perform a one-loop calculation using explicit
expressions of the propagators as warm-up exercises, and we will find
the Hall conductivity as well as the suppressed transverse conductivity.

\subsection{Real-time propagators}

We here adopt the real-time Schwinger-Keldysh formalism in the Keldysh
basis.  The propagators in the Keldysh basis and the standard ones on the
Schwinger-Keldysh paths (1,2) are transformed through the following
relations:
\begin{align}
  S^{f}_{RR} &:= \frac{1}{4} \bigl\langle(\psi_{f1} + \psi_{f2}) 
  (\bar{\psi}_{f1} + \bar{\psi}_{f2})\bigr\rangle 
  = \frac{1}{4}(S^{f}_{11}+S^{f}_{12}+S^{f}_{21}+S^{f}_{22})=-\ri S^{f}_K\,,
  \label{eq:SRRK}\\
  S^{f}_{RA} &:= \frac{1}{2} \bigl\langle(\psi_{f1} + \psi_{f2}) 
  (\bar{\psi}_{f1} - \bar{\psi}_{f2})\bigr\rangle 
  = \frac{1}{2}(S^{f}_{11}-S^{f}_{12}+S^{f}_{21}-S^{f}_{22})=-\ri S^{f}_R\,,\\
  S^{f}_{AR} &:= \frac{1}{2} \bigl\langle(\psi_{f1} - \psi_{f2}) 
  (\bar{\psi}_{f1} + \bar{\psi}_{f2})\bigr\rangle 
  = \frac{1}{2}(S^{f}_{11}+S^{f}_{12}-S^{f}_{21}-S^{f}_{22})=-\ri S^{f}_A\,,\\
  S^{f}_{AA} &:= \bigl\langle(\psi_{f1} - \psi_{f2}) 
  (\bar{\psi}_{f1} - \bar{\psi}_{f2})\bigr\rangle 
  =(S^{f}_{11}-S^{f}_{12}-S^{f}_{21}+S^{f}_{22}) = 0\,. 
\end{align}
The last identify generally holds, which is understood from the 
spectral representation of the propagators.  Now, to proceed further, 
we need expressions for the free propagators under the magnetic field
$B$.  The retarded/advanced propagator is given by a sum over the
Landau levels labeled by $n$ as
\begin{equation}
  S^{f}_{R/A}(p) = \sum_{n=0}^\infty 
  \frac{-S^{f}_n(p)}{p_0^2-\varepsilon_{f n}^2 \pm \ri \epsilon p_0}
  = \sum_{n=0}^\infty 
  \frac{-S^{f}_n(p)}{p_\parallel^2-m_{f n}^2 \pm \ri \epsilon p_0}\,,
  \label{eq:SRA}
\end{equation}
where $R$ corresponds to $+$ and $A$ to $-$ in front of
$\ri\epsilon p_0$.   The Landau quantized energy dispersion is 
$\varepsilon_{f n}=\sqrt{p_z^2+2|q_{f}B|n+m_{f}^2}$.  To go to the
last expression above, we defined $m_{f n}^2:=2|q_{f}B| n+m_{f}^2$
and the transverse and the longitudinal momenta,
$p^\mu_\perp=(0,p_x,p_y,0)$ and 
$p^\mu_\parallel=(p_0,0,0,p_z)$, in accord to the convention with
$\bB$ chosen along the $z$ axis.

The numerator $S^{f}_n(p)$ has Dirac index structures decomposed as 
\begin{equation}
  \label{eq:AB}
  S^{f}_{n}(p) = (\Slash{p}_\parallel+m_f) \bigl[
  P^{f}_+ A_{n+}(4\xi^{f}_p) + P^{f}_- A_{n-}(4\xi^{f}_{p}) \bigr]
  + \Slash{p}_\perp B_n(4\xi^{f}_{p}) 
\end{equation}
with $\xi^{f}_p:=|\bp_\perp|^2/(2|q_{f}B|)$ and 
\begin{align}
  \label{eq:An+}
  A_{n+}(\xi) &:= 2\rme^{-\xi/2} (-1)^n L_{n}(\xi)\,,\\
  \label{eq:An-}
  A_{n-}(\xi) &:= 2\rme^{-\xi/2} (-1)^{n-1} L_{n-1}(\xi)\,,\\
  \label{eq:Bn}
  B_n(\xi)  &:= 4\rme^{-\xi/2} (-1)^{n-1} L_{n-1}^{(1)}(\xi)\,.
\end{align}
Here, $L_n(\xi)=L_n^{(0)}(\xi)$, and $L_n^{(\alpha)}(\xi)$ represents
the generalized Laguerre Polynomials~\cite{Gusynin:1994xp}.  The above
form is the same one as used in Ref.~\cite{Fukushima:2012kc}.  For the
propagator and related integrals, see also
Refs.~\cite{Hattori:2012je,Hattori:2012ny}.  In the above expression
$P^{f}_{\pm}$ represents the projection operator defined by
$P^{f}_\pm:=\frac{1}{2}[1\pm \sgn(q_{f}B)\,\ri\gamma^1\gamma^2]$.  For
convenience to treat the projection operators, let us introduce,
\begin{equation}
  \gamma^\pm := \frac{1}{2}(\gamma^1\pm \ri \gamma^2)\,,
  \label{eq:gammapm}
\end{equation}
which satisfy $(\gamma^\pm)^2=0$ and
$\gamma^\pm \gamma^\mp=-(1\pm \ri \gamma^{1}\gamma^{2})/2$, with which
the projection operators can be represented as
$P^{f}_{\pm \sgn(q_{f}B)} = -\gamma^\pm \gamma^\mp$ and we see
$\gamma^\pm P^{f}_{\mp\sgn(q_{f}B)}=\gamma^\pm$
follows.  In our later calculations we will make use of these relations. 

We note that the propagators must satisfy the fluctuation dissipation 
relation, which is expressed as 
\begin{equation}
  \label{eq:SK}
  S^{f}_K(p) = \biggl[ \frac{1}{2}-n_F(p_0-\mu_{f}) \biggr]
  \bigl[ S_{R}^{f}(p) - S_A^{f}(p) \bigr]
  = \ri\biggl[ \frac{1}{2} - n_F(p_0-\mu_{f}) \biggr] \rho_{f}(p)\,. 
\end{equation}
Here, $n_F(p_0-\mu_f)$ is the Fermi-Dirac distribution function with
an argument shifted by a chemical potential $\mu_f$ for $f$-flavor
quark.  The spectral function is defined by 
$\ri \rho_{f}(p):=S_{R}^{f}(p)-S_{A}^{f}(p)$, which is, for the
free propagator,
\begin{equation}
  \rho_{f}(p) = \sum_{n=0}^\infty 
  S^{f}_n(p) (2\pi)\sgn(p_0) \delta(p_0^2-\varepsilon_{f n}^2)\,,
  \label{eq:spectral}
\end{equation}
as easily derived from Eq.~\eqref{eq:SRA}.

\subsection{Loop integrals}

We shall start with the simplest loop integral, that is, the electric
charge density, $n_e:=\langle j^0\rangle$, for a non-interacting Fermi
gas exposed to the magnetic field.  In terms of the propagator we
write it as
\begin{align}
  n_e &= -\sum_f q_f\int \frac{\rd^4 p}{(2\pi)^4}\;
  \tr\bigl[\gamma^0 S^f_{RR}(p)\bigr] \notag\\
  &= -\Nc \sum_f q_f \sum_{n=0}^\infty \int\frac{\rd^2 p_\parallel}{(2\pi)^2}
  \biggl[ \frac{1}{2} - n_F(p_0\!-\!\mu_f) \biggr]
  (2\pi)\sgn(p_0) \delta(p_0^2-\varepsilon_{f n}^2)
  \int\frac{\rd^2 p_\perp}{(2\pi)^2}\; \tr\bigl[\gamma^0 S^{f}_n(p)\bigr]\,,
\end{align}
which follows from Eqs.~\eqref{eq:SRRK}, \eqref{eq:SK}, and
\eqref{eq:spectral}.
The overall minus sign comes from the fermion loop.  We can easily
take the Dirac trace to find,
\begin{equation}
  \label{eq:twoLag}
  \tr\bigl[\gamma^0 S^f_n(p)\bigr] =
  4 e^{-4\xi^f_p/2}(-1)^n p_0\Bigl[L_n(4\xi_p^f)-L_{n-1}(4\xi_p^f)\Bigr]\,.
\end{equation}
The transverse integral of the Laguerre Polynomial counts the Landau
degeneracy factor as seen from
\begin{equation}
  \int\frac{\rd^2 p_\perp}{(2\pi)^2}\; \rme^{-4\xi_p^f/2} (-1)^n
  L_n(4\xi_p^f) = \frac{|q_f B|}{2\pi} \int_0^\infty d\xi_p^f\;
  \rme^{-4\xi_p^f/2} (-1)^n L_n(4\xi_p^f) = \frac{|q_f B|}{4\pi}
\label{eq:trans_integ}
\end{equation}
for $n\geq0$.  Using the above integral expression, we can simplify
the electric charge density into the following form:
\begin{align}
  n_e &= - \Nc \sum_f  \sum_{n=0}^\infty \alpha_n\, \frac{q_f|q_f B|}{\pi}
  \int\frac{\rd^2 p_\parallel}{(2\pi)^2}\; p_0
  \biggl[ \frac{1}{2}-n_F(p_0-\mu_f) \biggr]
  (2\pi)\sgn(p_0) \delta(p_0^2-\varepsilon_{f n}^2) \notag\\
  &= \Nc \sum_f \sum_{n=0}^\infty \alpha_n\, \frac{q_f|q_f B|}{2\pi}
  \int \frac{\rd p_z}{2\pi}\Bigl[ n_F(\varepsilon_{f n}-\mu_f)
    -n_F(\varepsilon_{f n}+\mu_f) \Bigr]\,.
\end{align}
This is exactly the expression obtained by standard procedures to
replace the phase space integration with the Landau level sum.  Here,
as usual, we introduced the spin degeneracy factor $\alpha_n$ defined by
\begin{equation}
  \alpha_n = \begin{cases}
  1 & (n=0) \\ 2 & (n>0)
  \end{cases}
\end{equation}
appearing from two Laguerre polynomials in Eq.~\eqref{eq:twoLag}.
This factor takes care of the fact that the LLL at $n=0$ has only one
spin state.

We are now ready for going into the next exercise of loop integrals.
In the Keldysh basis, we can write the conductivity at the one loop level or
the polarization tensor as
\begin{equation}
  \begin{split}
  \Pi_R^{\mu\nu}(k) &= \sum_f (-1)\,\ri q_f^2\int\frac{\rd^4 p}{(2\pi)^4}
  \tr\bigl[ \gamma^\mu S^f_{RR}(k+p) \gamma^\nu S^f_{AR}(p) \bigr] \\
  &\qquad + \sum_f (-1)\,\ri q_f^2 \int\frac{\rd^4 p}{(2\pi)^4}
  \tr\bigl[ \gamma^\mu S^f_{RA}(k+p) \gamma^\nu S^f_{RR}(p) \bigr]\,.
\end{split}
\label{eq:Pioneloop}
\end{equation}
For transport coefficients we need to evaluate the above polarization
tensor in the $\bk\to \bzero$ limit.  For this, we first take the
$\bk_\perp\to \bzero$ limit, under which the trace with the propagator
numerators leads to
\begin{align}
  \label{eq:SS}
  & \lim_{\bk_\perp\to\bzero}\tr\bigl[\gamma^- \Dn^f_n(p+k)\gamma^+ \Dn^f_m(p)\bigl] \notag\\
  & = 4\, \rme^{-4\xi_p^{f}} (-1)^{n-m}\;\tr\bigl\{ \gamma^- \bigl[
  (\Slash{p}_\parallel+\Slash{k}_\parallel+m_{f})
  (P^{f}_+ L_n(4\xi_p^{f}) - P^{f}_- L_{n-1}(4\xi_p^{f}))
  - 2\Slash{p}_\perp L_{n-1}^{(1)}(4\xi_p^{f}) \bigr] \notag\\
  &\qquad \times \gamma^+ \bigl[
  (\Slash{p}_\parallel+m_{f})(P^{f}_+ L_m(4\xi_p^{f}) - P^{f}_- L_{m-1}(4\xi_p^{f}))
  - 2\Slash{p}_\perp L_{m-1}^{(1)}(4\xi_p^{f}) \bigr] \bigr\} \notag\\
    &= 8\, \rme^{-4\xi_p^{f}} (-1)^{n-m+1}\;
  \bigl[(p_\parallel+k_\parallel)\cdot p_\parallel - m_{f}^2\bigr]\notag\\
  &\qquad\times \bigl[\theta(q_{f}B)L_n(4\xi_p^{f})L_{m-1}(4\xi_p^{f}) +
    \theta(-q_{f}B) L_{n-1}(4\xi_p^{f})L_m(4\xi_p^{f}) \bigr] \,,
\end{align}
where we used $\gamma^\pm$ of Eq.~\eqref{eq:gammapm} instead of the
coordinate basis.
We note that we used
$\tr\bigl[\gamma^- \Slash{p}_\perp\gamma^+ \Slash{p}_\perp\bigr]=0$ to
arrive at the last expression.  Then, we can immediately take the
Dirac trace and perform the transverse momentum integration to get
\begin{align}
  & \lim_{\bk_\perp\to\bzero} \int\frac{\rd^2 p_\perp}{(2\pi)^2}\;
  \tr\bigl[ \gamma^- \Dn^{f}_n(p+k)\gamma^+ \Dn^{f}_m(p) \bigr] \notag\\
  &= 8\, \bigl[ (p_\parallel+k_\parallel)\cdot p_\parallel-m_f^2\bigr]
    \, \frac{|q_f B|}{8\pi}\,\bigl[\theta(q_{f}B)\delta_{n,m-1}
    +\theta(-q_{f}B)\delta_{n-1,m}\bigr]\,.
\end{align}
From the above expression it is straightforward to read the
conductivity.  For this purpose to identify the Hall and the
transverse conductivities it is useful to express physical quantities
in terms of $\pm$ coordinates in accord to $\gamma^\pm$, that is,
\begin{equation}
  j^\pm  := \frac{1}{2}(j^1 \pm \ri j^2)\,, \qquad
 \sigma^{-+} := \frac{1}{2}(\sigma_\perp+\ri \sigma_H)\,. 
\end{equation}
We can readily confirm that other components (apart from the 
longitudinal ones) are irrelevant, i.e., $\sigma^{++}=\sigma^{--}=0$. 
Then, we do not have to compute $\sigma_H$ and $\sigma_\perp$
individually but we can just separate them from the real part and the
imaginary part of $\sigma^{-+}$.  We also use a trick to simplify the
calculation through the following relation:
\begin{equation}
  \sigma^{-+} = \lim_{k_0\to0} \lim_{\bk\to\bzero}\,
  \frac{1}{2\ri k_0}\; \bigl[ \Pi_R^{-+}(k) - \Pi_A^{-+}(k) \bigr]
  = \lim_{k_0\to0} \lim_{\bk\to \bzero}\,
  \frac{1}{\ri k_0}\; \Pi_R^{-+}(k)\,.
  \label{eq:trick}
\end{equation}
This relation holds in our present calculation, but may not be exact
in general.  In the present case, thanks to the above relation, we can
simplify the algebra and little more calculational steps eventually
lead us to
\begin{align}
  \sigma^{-+} &= \lim_{k_0\to0}\lim_{\bk\to\bzero}\;
  \frac{1}{\ri k_0}\,\Pi_R^{-+}(k) \notag\\
  &= \Nc\sum_f q_f^2\lim_{k_0\to0}\lim_{k_z\to 0}\;
  \frac{1}{\ri k_0}\sum_{n,m}
  \int \frac{\rd^4 p}{(2\pi)^4}\;\tr\bigl[
  \gamma^- \Dn^{f}_n(k+p) \gamma^+ \Dn^{f}_m(p)\bigr] \notag\\
  &\quad \times \Biggl\{ \biggl[ \frac{1}{2}-n_F(p_0+k_0-\mu_{f})\biggr]
  (2\pi)\sgn(p_0+k_0)\delta\bigl((p_\parallel+k_\parallel)^2-m_{f n}^2\bigr)
  \,\frac{1}{p_\parallel^2-m_{f m}^2-\ri\epsilon p_0} \notag\\
  &\qquad +\frac{1}{(p_\parallel+k_\parallel)^2-m_{f n}^2+\ri\epsilon(p_0+k_0)}
  \biggl[ \frac{1}{2}-n_F(p_0-\mu_{f})\biggr]
  (2\pi)\sgn(p_0)\delta(p_\parallel^2-m_{f m}^2) \Biggr\} \notag\\
  &= -\ri\Nc\sum_f \sgn(q_{f}B)\;\frac{q_f^2}{2\pi}\sum_n \alpha_n 
  \int \frac{\rd^2 p_\parallel}{(2\pi)^2}\;p_0\biggl[
  \frac{1}{2}-n_F(p_0-\mu_{f})\biggr]
  (2\pi)\sgn(p_0)\delta(p_\parallel^2-m_{f n}^2) \notag\\
  &=\ri \Nc\sum_f \sgn(q_{f}B)\;\frac{q_f^2}{4\pi}\sum_n \alpha_n \int
    \frac{\rd p_z}{2\pi}
  \Bigl[ n_F(\varepsilon_{f n}\!-\!\mu_{f}) - n_F(\varepsilon_{f n}\!+\!\mu_{f}) \Bigr]
  =\frac{\ri}{2}\cdot \frac{n_e}{B}\,.
\end{align}
The real part and the imaginary part result in $\sigma_\perp=0$ and
$\sigma_H=n_e/B$, respectively, as advertised in Eq.~\eqref{eq:Hall}.
To find a nonzero value of $\sigma_\perp$ we need to go to the two loop
calculation which is beyond our present scope.  Here, we just give a
parametric estimate, that is,
\begin{equation}
  \frac{\sigma_\perp}{T} \;\sim\; \frac{g^2T^2}{|qB|}\,,
\end{equation}
which is small in our condition of $\sqrt{|qB|}\gg gT$.

This parametric form can be understood from one self-energy insertion
of $\Sigma$ to one of the fermion propagators, i.e., in the two-loop
order the left-hand side of Eq.~\eqref{eq:SS} should be replaced with
\begin{equation}
  \lim_{\bk\to\bzero}\tr\bigl[\gamma^- \Dn^{f}_{n}(p+k)\gamma^+ \Dn^{f}_m(p)
  \Sigma^{f}(p)\Dn^{f}_l(p)\bigr]\,.
\end{equation}
The leading behavior of the self-energy is $\sim g^2 T^2$, while the
propagator is of order $1/\Delta\varepsilon \sim T/|q B|$ where
$\Delta\varepsilon$ is an energy gap associated with adjacent Landau
levels.  Thus, the combination of these factors leads to
$\sigma_\perp \sim g^2 T^2\cdot T/|q B| = g^2 T^2/|q B|$, which
explains Eq.~\eqref{eq:sigmaperp}.

\section{Longitudinal conductivity}
\label{sec:longitudinal}

\begin{figure}
  \centering 
  \includegraphics[width=0.35\textwidth]{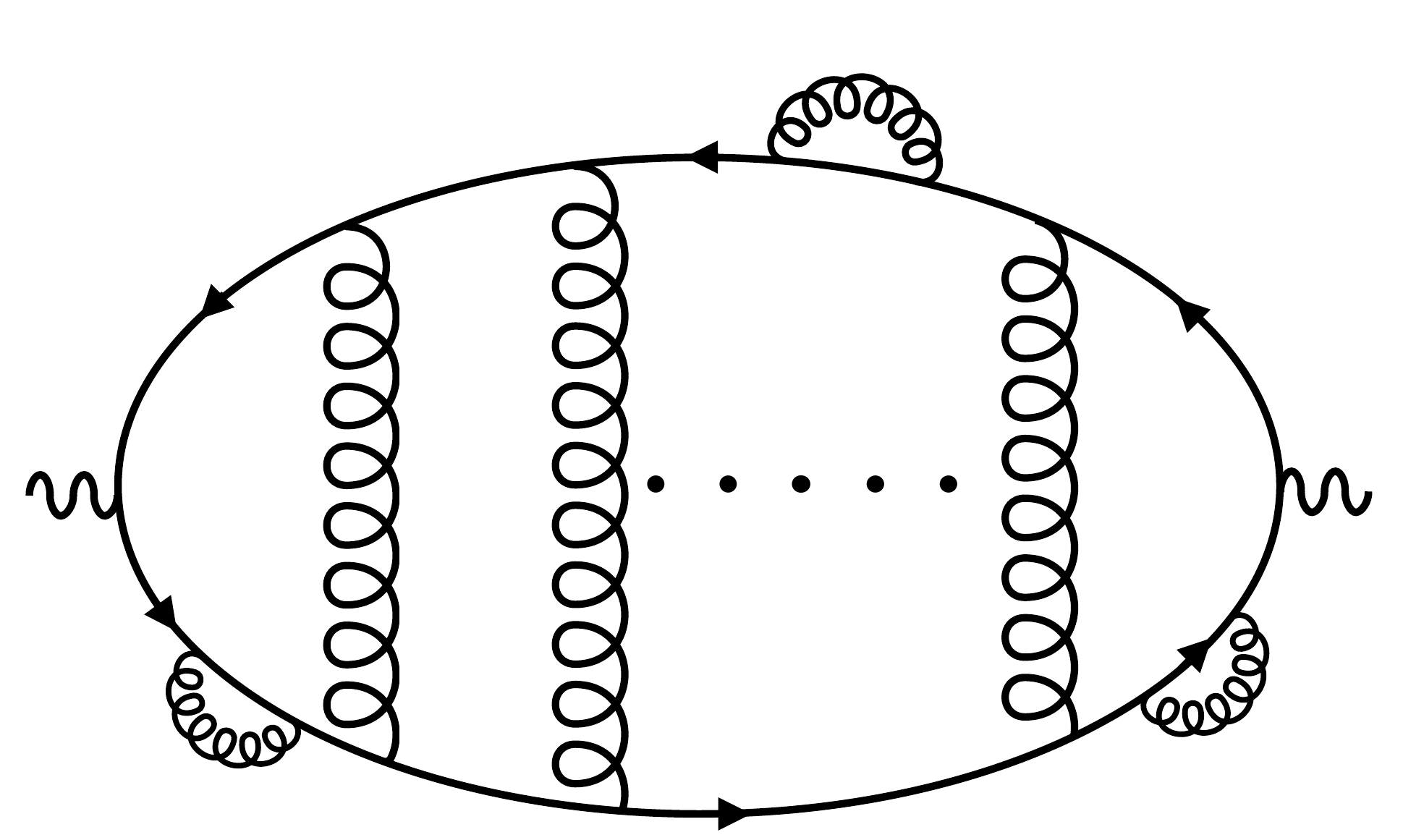}
  \caption{Typical diagram contributing to the leading order 
  calculation of the longitudinal conductivity.}
  \label{fig:typical}
\end{figure}

\begin{figure}
  \centering
  \parbox[c]{0.45\textwidth}{\includegraphics[width=0.45\textwidth]{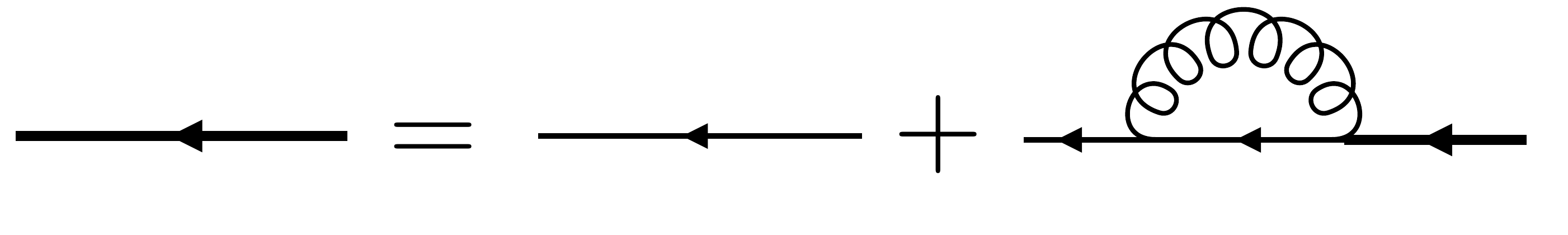}}
  \qquad\qquad
  \parbox[c]{0.4\textwidth}{\includegraphics[width=0.4\textwidth]{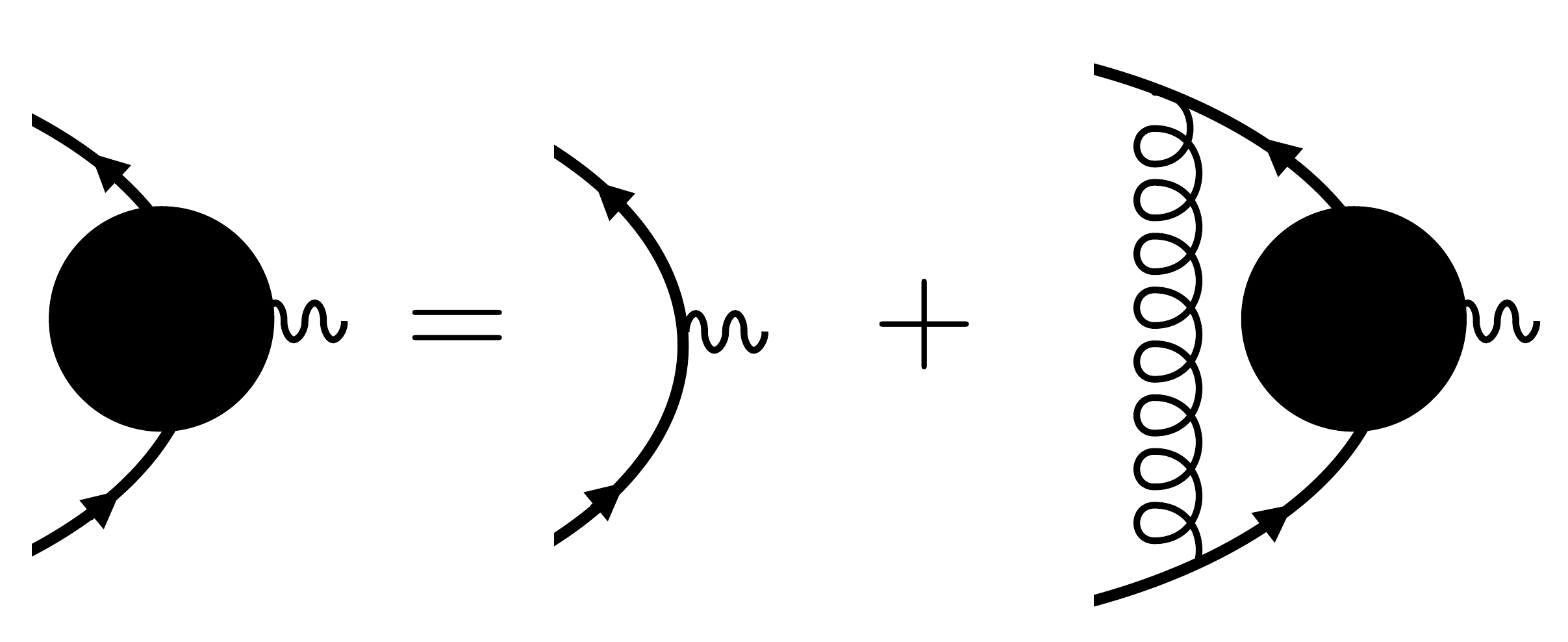}}
  \caption{Illustration of the Bethe-Salpeter equations;  the resummed
  propagator with self-energy insertions (left) and the resummed
  vertex with ladder diagrams (right).}
  \label{fig:BSequation}
\end{figure}

This section details the derivation of $\sigma_\parallel$.  To this
end, we should start with the Kubo formula as we did for $\sigma^{-+}$
previously.  The calculation for $\sigma_\parallel$ is technically
involved and we must take account of the resummation of pinching
singularities, which is a common technique used also for other transport
coefficient calculations (see Ref.~\cite{Hidaka:2010gh} for example).
The pinching singularities generally appear from the following type of
integral:
\begin{equation}
  \int \frac{\rd k_0}{(2\pi)}\; F(k_0) \cdot
  \frac{1}{k_0-\varepsilon +\ri \gamma} \cdot
  \frac{1}{k_0-\varepsilon -\ri \gamma} \;\sim\;
  \frac{F(\varepsilon)}{2\gamma}\,.
\end{equation}
A typical contribution to the leading order calculation for 
$\sigma_\parallel$ is diagrammatically shown in 
Fig.~\ref{fig:typical}.  In Sec.~\ref{sec:diagram} we will spell out
all field-theoretical transformations, and in this section, let us
postpone the derivation and just adopt a kinetic description as a
result of the resummation.  Although it is nontrivial to prove
the equivalence as addressed in Sec.~\ref{sec:diagram}, the kinetic
equations provide us with the most efficient approach to evaluate
physical observables taking account of the resummation.  In fact, in
field-theoretical language, the kinetic equations are the
Bethe-Salpeter equations, as illustrated schematically in
Fig.~\ref{fig:BSequation}.  We can easily confirm
that a diagram like Fig.~\ref{fig:typical} is produced by the
self-energy insertion and the resummed vertex formulated by
iterative processes in Fig.~\ref{fig:BSequation}.

We briefly summarize our strategy to reach $\sigma_\parallel$.  We
will write down the kinetic or the linearized Boltzmann equations with
the collision terms at the lowest scattering order.  Then, we will
compute a distortion, $\delta f$, on the Fermi-Dirac distribution
function, $n_F$, in response to the applied electric field, which
needs an operator inversion on the Boltzmann equations.
Once $\delta f$ is worked out in this way, it is easy to represent the
electric current in a form proportional to the electric field, and
the proportionality coefficient is nothing but the electric conductivity.

\subsection{Boltzmann equations and the formal solution}
\label{sec:Boltzmann}

The Bethe-Salpeter equations can be translated to the linearized
Boltzmann equations with the collision term of scattering processes.
The corresponding Boltzmann equations for quarks, anti-quarks, and
gluons take the following forms, respectively,
\begin{equation}
  \label{eq:BoltzmannEq}
  \begin{split}
  & 2\Pmu_p^\mu \bigl(\partial_\mu
  + q_{f} F_{\nu\mu} \partial_{p_\nu} \bigr) f_p = -C[f]\,,\\
  & 2\bar{\Pmu}_{p'}^\mu \bigl(\partial_\mu
  - q_{f} F_{\nu\mu} \partial_{p'_\nu} \bigr) \bar{f}_{p'} = -\bar{C}[f]\,,\\
  & 2k^\mu \partial_\mu g_k = -\tilde{C}[f]\,,
  \end{split}
\end{equation}
where $\partial_{p_\nu}:=\partial/\partial p_\nu$ and $C[f]$,
$\bar{C}[f]$, and $\tilde{C}[f]$ represent the collision terms.  We
introduced notations, $2\Pmu_p^\mu := \bar{u}(p)\gamma^\mu u(p)$ and
$2\bar{\Pmu}_{p'}^\mu := \bar{v}(p')\gamma^\mu v(p')$, with the spinor
wave functions $u(p)$ and $v(p')$ for particle and anti-particle, respectively
[the precise definition of the wave functions will be given in
Eq.~\eqref{eq:waveFunctions}].
For these expressions we use a sloppy notation for the
indices;  the subscript $p$, $p'$, and $k$ represent not only the
momenta but also the Landau level $n$, the angular momentum $l$, the
spin $s$, the color $c$, and the flavor $f$.

To solve the Boltzmann equation perturbatively, we expand the
distribution functions around the thermal equilibrium, that is,
\begin{equation}
  f_p = \feq(p) + \delta f_p\,,\quad
  \bar{f}_{p'} = \fbareq(p') + \delta \bar{f}_{p'}\,,\quad
  g_k = \gequiv(k) + \delta g_k
\end{equation}
with $\feq(p):=n_{F}(\varepsilon_{fn}-\mu_{f})$,
$\fbareq(p):=n_{F}(\varepsilon_{fn}+\mu_{f})$, and
$\gequiv(k):=n_{B}(\omega_{k})$ at the local rest frame, where
$\omega_k=|\bk|$ is the energy of gluons.  In the present problem we
are interested in the longitudinal conductivity, so we can consider a
homogeneous electric field $E_z$ along the $z$ axis up to the linear
order in terms of $E_z$ and $\delta f_p$, $\delta\bar{f}_{p'}$,
$\delta g_k \,\propto E_z$.  We note that we consider only the
``diagonal components'' of the distribution functions in spin space,
but for our purpose especially without the axial charge coupling, this
treatment is sufficient as long as our computation is closed in the
linear response regime.  For some recent discussions on the full spin
dependent distribution functions, see
Refs.~\cite{Weickgenannt:2019dks,Gao:2019znl,Hattori:2019ahi,Wang:2019moi}.

Let us first look at the left-hand side of the Boltzmann
equations~\eqref{eq:BoltzmannEq}.  Since the derivative and the field
strength are already of order of $E_z$, we can safely drop higher
order terms involving
$\delta f_p$, $\delta\bar{f}_{p'}$, $\delta g_k$ and
substitute $\feq(p)$, $\fbareq(p')$, and $\gequiv(k)$ for $f_p$,
$\bar{f}_{p'}$, and $g_k$.  Our assumption of homogeneity makes the
spatial derivatives vanishing, and we eventually find,
\begin{equation}
  \label{eq:BoltzmannEq2}
  \begin{split}
  & 2\Pmu_p^0 \,\bigl(\partial_0 + q_{f} E_z\,\partial_{p_z} \bigr) \feq(p)
  = -\beta W_p\Bigl( -p_z\partial_0 u_z
  + q_{f} E_z\frac{p_z}{\varepsilon_{f n}} \Bigr)\,,\\
  & 2\Pmu_{p'}^0 \,\bigl(\partial_0 - q_{f} E_z\,\partial_{p'_z} \bigr)\fbareq(p')
  = -\beta \bar{W}_{p'} \Bigl( -p'_z\partial_0 u_z
  -q_{f} E_z\frac{p'_z}{\varepsilon_{f n'}} \Bigr)\,,\\
  & 2\omega_k \,\partial_0 \gequiv(k)
    = -\beta \tilde{W}_k (-k_z\partial_0 u_z)\,,
  \end{split}
\end{equation}
where we introduced several new notations.  We factorize the
derivatives of the thermal distribution functions by introducing the
following functions:
\begin{align}
  W_p &:= 2P_p^0 \, \feq(p)[1-\feq(p)]\,,\\
  \bar{W}_{p'} &:= 2P_{p'}^0\, \fbareq(p')[1-\fbareq(p')]\,,\\
  \tilde{W}_k &:= 2\omega_k\, \gequiv(k)[1+\gequiv(k)]\,,
\end{align}
which will be used as the weight functions in the inner product in
later calculations.  The first terms appear from time-dependent fluid
velocity $u_z$ which is induced by the $E_z$ effect.  We note that in
the linear response regime no fluid velocity is developed yet, but
$\partial_0 u_z$ can be nonvanishing.  One can easily understand how
$\partial_0 u_z$ terms emerge by replacing
$\varepsilon_{fn}\to p_{fn}\cdot u$ and
$\omega_k\to k\cdot u$ with the fluid velocity $u^\mu$.  Now, we
should quantify $\partial_0 u_z$ in response to $E_z$; for this
purpose we can use the leading order hydrodynamic equation,
$\partial_0 u_z = n_e E_z/(\calE+\calP_z)$, which can be immediately
understood from $\partial_\mu T^{\mu\nu}=F^{\nu}_{~\mu}\, j_\text{em}^\mu$.  Here, as mentioned below Eq.~\eqref{eq:jsubt}, $\calE$ is the energy
density and $\calP_z$ is the pressure in the $z$ direction, which are
explicitly given, respectively, as
\begin{align}
  \calE &:= \Nc \sum_{f, n} \alpha_n\, \frac{|q_f B|}{2\pi}
  \int \frac{\rd p_z}{2\pi}\,\varepsilon_{f n}
  \bigl[ \feq(p)+\fbareq(p) \bigr]
  + 2(\Nc^2-1)\int\frac{\rd^3 k}{(2\pi)^3}\,\omega_k\,\gequiv(k)\,, \\
  \calP_z &:= \Nc \sum_{f, n} \alpha_n\, \frac{|q_f B|}{2\pi}
  \int \frac{\rd p_z}{2\pi}\frac{p_z^2}{\varepsilon_{f n}}
  \bigl[ \feq(p)+\fbareq(p) \bigr]
  +2(\Nc^2-1)\int\frac{\rd^3 k}{(2\pi)^3}\frac{k_z^2}{\omega_k}\gequiv(k) \,.
\end{align}
We have finished the preparation, and we can now specify the collision
integrals to solve $\delta f_p$, $\delta \bar{f}_{p'}$, and
$\delta g_k$.  Before doing so, however, let us introduce symbolic
representations to further sort out expressions algebraically.

Instead of the original Boltzmann equations~\eqref{eq:BoltzmannEq}, we
write them concisely as
\begin{equation}
  \label{eq:BoltzmannSymbol}
  \calS = \calL \chi\,.
\end{equation}
For convenience we move $W_p$, $\bar{W}_{p'}$, and $\tilde{W}_k$ as
well as $E_z$ from the left-hand side to the right-hand side, and then
we find,
\begin{equation}
  \calS := \calJ^z - \frac{ n_e \calT^{0z}}{\calE + \calP_z}
\label{eq:Sdef}
\end{equation}
from Eq.~\eqref{eq:BoltzmannEq2}, where
\begin{equation}
  \label{eq:J}
  \calJ^\mu := \begin{pmatrix}
  q_{f} p^\mu / \varepsilon_{f n} \\
  - q_{f} p'^\mu / \varepsilon_{f n'} \\
  0
  \end{pmatrix}\,,\qquad
  \calT^{0\mu} := \begin{pmatrix}
  p^\mu \\ p'^\mu \\ k^\mu
  \end{pmatrix}\,.
\end{equation}
Interestingly, this form is analogous to the expectation value of the
each flavor current with subtraction as in Eq.~\eqref{eq:jsubt}, which
will turn out to be crucial to make the calculation well-defined.  The
right-hand side of Eq.~\eqref{eq:BoltzmannSymbol}, with $W_p$,
$\bar{W}_{p'}$, $\tilde{W}_k$, and $E_z$ moved from the left-hand
side, simply reads,
\begin{equation}
  \calL \chi := \frac{1}{E_z}
  \begin{pmatrix}
   \displaystyle \frac{1}{\beta W_p }C[f] \\
   \displaystyle \frac{1}{\beta \bar{W}_{p'}}\bar{C}[f] \\
   \displaystyle \frac{1}{\beta \tilde{W}_k }\tilde{C}[f]
 \end{pmatrix}\,.
 \label{eq:Lchi}
\end{equation}
In the next subsection we will present all the details about the collision
terms.  For the moment it should be noted that $C=\bar{C}=\tilde{C}=0$
for equilibrium distributions, and so $\calL \chi$ should be of order
of $\delta f_p$, $\delta \bar{f}_{p'}$, and $\delta g_k$.

At this point, we shall introduce some more notations which are useful
to simplify actual computations.  To eliminate kinematic factors in
the denominators of Eq.~\eqref{eq:Lchi}, we rescale $\delta f_p$,
$\delta \bar{f}_{p'}$, and $\delta g_k$ as
\begin{equation}
  \label{eq:parametrization}
  \begin{split}
  \delta f_p &= \beta \feq(p)[1-\feq(p)]\,  E_z\, \chi_p\,, \\
  \delta \bar{f}_{p'} &= \beta \fbareq(p')[1-\fbareq(p')]\,
   E_z\, \bar{\chi}_{p'}\,,\\
  \delta g_k &= \beta \gequiv(k)[1+\gequiv(k)]\,
   E_z\, \tilde{\chi}_k\,.
  \end{split}
\end{equation}
Introducing a vector symbol,
\begin{equation}
  \chi := 
  \begin{pmatrix}
  \chi_p \\
  \bar{\chi}_{p'} \\
  \tilde{\chi}_k
  \end{pmatrix}\,,
\end{equation}
we can interpret Eq.~\eqref{eq:Lchi} as a linear operation of $\calL$
onto $\chi$, which means that the formal solution,
$\chi = \calL^{-1} \calS$, gives the longitudinal conductivity as
follows.  We can write down the electric current parallel to $B$ along
the $z$ axis and the longitudinal conductivity as
\begin{align}
  \sigma_\parallel &=
 \frac{ j_z}{E_z}  
  = \Nc \sum_{f}\frac{q_{f}|q_f B|}{2\pi}\sum_{n=0}^{\infty} \alpha_n
  \int\frac{\rd p_z}{2\pi}\, \frac{p_z}{\varepsilon_{f n}}
                    \biggl(\frac{\delta f_p}{E_z} - \frac{\delta \bar{f}_p}{E_z}\biggr) \notag\\
  &= \beta\Nc \sum_f \frac{q_f |q_{f} B|}{2\pi}
  \sum_{n=0}^\infty \alpha_n 
  \int\frac{\rd p_z}{2\pi}\,\frac{p_z}{\varepsilon_{f n}}\Bigl\{
  \feq(p)[1-\feq(p)]\chi_p 
  - \fbareq(p)[1-\fbareq(p)]\bar{\chi}_p\Bigr\}\,. 
  \label{eq:cond}
\end{align}
Up to now, all necessary ingredients except for the collision terms
have been presented.  However, these are not yet adequate for real
calculations; $\calL$ contains zero eigenvalues with the
eigenvectors, $\calC^a = \{\calJ^0 ,\calT^{0\mu} \}$ (i.e.,
hydrodynamic modes), corresponding to the charge and the
energy-momentum conservations.  
Here, we suppress the flavor index below to avoid cumbersome notations.

To make $\calL^{-1}$ well-defined, we should get rid of such zero
modes.  We must emphasize that this removal of zero modes is
\textit{not} an \textit{ad hoc} procedure;  $\calS = \calL\chi$ is a
perfectly well-defined equation to solve $\chi$ because of the special
structure of $\calS$.  To see this point in a clear way, our symbolic
representation is useful.

Since we want to discuss the projection operation, we need to define
an inner product.  A natural choice of the inner product of two
functions,
$A=(a_p,\bar{a}_{p'},\tilde{a}_{k})$ and 
$B=(b_p,\bar{b}_{p'},\tilde{b}_{k})$, should be
\begin{equation}
  \label{eq:inner}
  (A, B) := \int_p\, W_p\, a_p b_p 
  + \int_{p'} \bar{W}_{p'}\, \bar{a}_{p'} \bar{b}_{p'}
  + \int_k\, \tilde{W}_k\, \tilde{a}_k \tilde{b}_k \,,
\end{equation}
where we used a simplified notation for all the phase space sum,
\begin{equation}
  \int_p := \sum_{n,l,c,s,f} \int\frac{\rd^{3} p}{(2\pi)^{3}}\,
  \frac{1}{2\varepsilon_{f n}}\,,\qquad
    \int_k := \sum_{c,s} \int\frac{\rd^{3} k}{(2\pi)^{3}}\,
  \frac{1}{2\omega_{k}}\,. 
\end{equation}
With this definition of the inner product the longitudinal electric
conductivity~\eqref{eq:cond} takes an extremely simple form as
\begin{equation}
  \label{eq:ConductivityFormula}
  \sigma_\parallel =\beta(\calJ^z, \chi)
\end{equation}
from an explicit expression of $\calJ^z$ in Eq.~\eqref{eq:J}.  Now,
using the zero eigenvector $\calC$ and the inner product as defined
above, we introduce a projection operator $\calQ$ onto functional
space excluding zero eigenvalues as
\begin{equation}
  \calQ O := O - \sum_{a,b} \calC^a (\calC,\calC)^{-1}_{ab}
  (\calC^b, O) \,,
\end{equation}
where $(\calC,\calC)^{-1}_{ab}$ is the inverse matrix of
$(\calC^a,\calC^b)$.  We immediately see $\calQ^2 = \calQ$ and
$\calQ\calC^a=0$ by construction.  Using
an alternative expression of the charge density and the enthalpy,
$n_e = \beta(\calT^{0z}, \calJ^z)$ and
$\calE+\calP_z = \beta(\calT^{0z},\calT^{0z})$~\cite{Minami:2012hs},
we can rewrite Eq.~\eqref{eq:Sdef} into $\calS = \calQ\calJ^z$.
Because zero eigenvalues simply give zero, $\calL=\calL\calQ$
trivially follows, and the equation, $\calL\chi=\calS$, is equivalent to
$\calQ \calL \calQ \chi = \calQ\calS = \calS$, which can be solved as
$\chi = \calQ\calL^{-1}\calQ\calS$.
We eventually obtain,
\begin{equation}
  \label{eq:sigmazz}
  \sigma_\parallel = \beta(\calJ^z, \calQ\calL^{-1}\calQ\calS)
  = \beta(\calS, \calL^{-1} \calS)
  = \beta(\calS, \chi)\,.
 \end{equation}

\subsection{Collision terms}
\label{sec:collision}

The last missing pieces are the collision terms, $C[f]$, $\bar{C}[f]$,
and $\tilde{C}[f]$.  In this work we consider the weak coupling
expansion with $g^2 \ll 1$, and the lowest order contributions then
arise from $1\leftrightarrow 2$ processes.  We note that the typical
scale of $1\leftrightarrow 2$ processes is $\sim g^2 q_f B/T^2$ which
is much larger than the typical scale $\sim g^4$ of $2\leftrightarrow
2$ processes under our hierarchy~\eqref{eq:hierarchy}.

\begin{figure}
  \begin{minipage}[b]{0.16\linewidth}
  \centering 
    \includegraphics[width=0.6\linewidth]{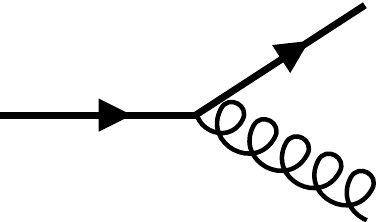}\\
    (a1)$q\to qg$
    \end{minipage}
  \begin{minipage}[b]{0.16\linewidth}
  \centering 
  \includegraphics[width=0.6\linewidth]{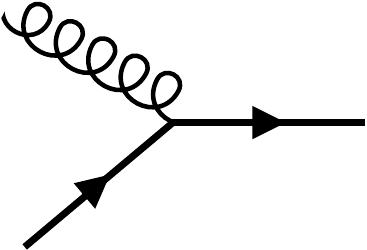} \\
   (a2) $qg\to q$
    \end{minipage}
  \begin{minipage}[b]{0.16\linewidth}
  \centering 
  \includegraphics[width=0.6\linewidth]{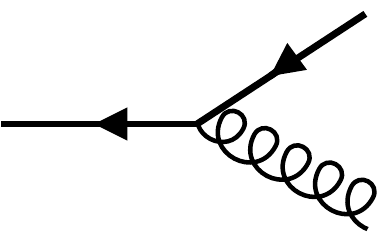}\\
   (b1) $\bar{q}\to \bar{q}g$
    \end{minipage}
  \begin{minipage}[b]{0.16\linewidth}
  \centering 
  \includegraphics[width=0.6\linewidth]{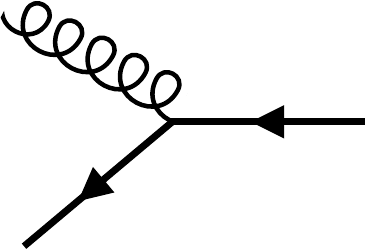}\\
  (b2) $\bar{q}g\to \bar{q}$
    \end{minipage}
  \begin{minipage}[b]{0.16\linewidth}
  \centering 
  \includegraphics[width=0.6\linewidth]{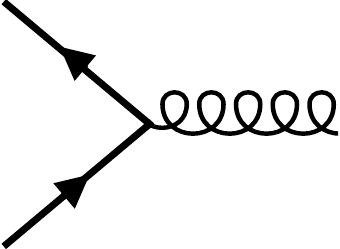} \\
   (c1) $q\bar{q}\to g$
    \end{minipage}
  \begin{minipage}[b]{0.16\linewidth}
  \centering 
  \includegraphics[width=0.6\linewidth]{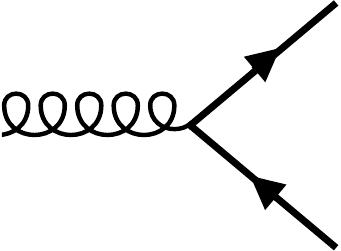} \\
  (c2) $g\to q\bar{q}$
    \end{minipage}
  \caption{Diagrams of the radiation process with a quark (a1) and a anti-quark (b1), 
  the pair annihilation (c1), and their inverse processes, (a2), (b2), and (c2).}
  \label{fig:rad}
\end{figure}

Such $1\leftrightarrow 2$ processes can be decomposed as
\begin{equation}
  \begin{split}
  C[f] &= C_{q\to qg}[f] + C_{qg\to q}[f] + C_{q\bar{q}\to g}[f]\,,\\
  \bar{C}[f] &= \bar{C}_{\bar{q}\to \bar{q}g}[f] + \bar{C}_{\bar{q}g\to \bar{q}}[f]
  + \bar{C}_{q\bar{q}\to g}[f]\,,\\
  \tilde{C}[f] &= \tilde{C}_{g\to q\bar{q}}[f] +\tilde{C}_{qg\to q}[f]
  + \tilde{C}_{\bar{q}g\to\bar{q}}[f] \,,
  \end{split}
\end{equation}
where the subscripts represent $1\leftrightarrow 2$ processes as
illustrated in Fig.~\ref{fig:rad}.  Once the scattering amplitude
$\calM$ is given (which we will compute later), the kinematical
assignments of the distribution functions lead to the following
expressions:
\begin{align}
  C_{q\to qg}[f] &= \int_{k,p'} |\calM_{p\to p'+k}|^2
  (2\pi)^4\delta^{(4)}(p\!-\!k\!-\!p')
  \Bigl[ f_p(1 \!-\! f_{p'})(1\! +\! g_k) - (1 \!-\! f_p)f_{p'} g_k \Bigr],\\
  C_{qg\to q}[f] &= \int_{k,p'} |\calM_{p+k\to p'}|^2
  (2\pi)^4\delta^{(4)}(p\!+\!k\!-\!p')
  \Bigl[ f_p g_k(1 \!-\! f_{p'}) - (1 \!-\! f_p)(1 \!+\! g_k)f_{p'} \Bigr],\\
  C_{q\bar{q}\to g}[f] &= \int_{k,p'} |\calM_{p+p'\to k}|^2
  (2\pi)^4\delta^{(4)}(p\!+\!p'\!-\!k)
  \Bigl[ f_p\bar{f}_{p'}(1 \!+\! g_k) - (1 \!-\! f_p)(1 \!-\! \bar{f}_{p'})g_k \Bigr]
\end{align}
for quarks and
\begin{align}
  \bar{C}_{\bar{q}\to\bar{q}g}[f] &= \int_{k,p'} |\calM_{p\to p'+k}|^2
  (2\pi)^4\delta^{(4)}(p\!-\!k\!-\!p')
  \Bigl[ \bar{f}_p(1 \!-\! \bar{f}_{p'})(1 \!+\! g_k)
  - (1 \!-\! \bar{f}_p)\bar{f}_{p'}g_k \Bigr],\\
  \bar{C}_{\bar{q}g\to \bar{q}}[f] &= \int_{k,p'} |\calM_{p+k\to p'}|^2
  (2\pi)^4\delta^{(4)}(p\!+\!k\!-\!p')
  \Bigl[ \bar{f}_p g_k(1 \!-\! \bar{f}_{p'})
  -( 1 \!-\! \bar{f}_p)(1 \!+\! g_k)\bar{f}_{p'} \Bigr],\\
  \bar{C}_{q\bar{q}\to g}[f] &= \int_{k,p'} |\calM_{p+p'\to k}|^2
  (2\pi)^4\delta^{(4)}(p\!+\!p'\!-\!k)
  \Bigl[ \bar{f}_pf_{p'}(1 \!+\! g_k) - (1 \!-\! \bar{f}_p)(1 \!-\! f_{p'})g_k \Bigr]
\end{align}
for anti-quarks and
\begin{align}
  \tilde{C}_{g\to q\bar{q}}[f] &= \int_{p,p'}|\calM_{k\to p+p'}|^2
  (2\pi)^4\delta^{(4)}(k\!-\!p\!-\!p') \Bigl[
  g_k(1 \!-\! f_p)(1 \!-\! \bar{f}_{p'})
  - (1 \!+\! g_k)f_p \bar{f}_{p'} \Bigr],\\
  \tilde{C}_{qg\to q}[f] &= \int_{p,p'} |\calM_{k+p\to p'}|^2
  (2\pi)^4\delta^{(4)}(k\!+\!p\!-\!p') \Bigl[
  g_kf_{p}(1 \!-\! f_{p'}) - (1 \!+\! g_k)(1 \!-\! f_{p})f_{p'}  \Bigr],\\
  \tilde{C}_{\bar{q}g\to \bar{q}}[f] &= \int_{p,p'} |\calM_{k+p'\to p}|^2
  (2\pi)^4\delta^{(4)}(k\!+\!p'\!-\!p) \Bigl[
  g_k\bar{f}_{p'}(1 \!-\! \bar{f}_p)
  -(1 \!+\! g_k)\bar{f}_p(1 \!-\! \bar{f}_{p'}) \Bigr]
\end{align}
for gluons.  Here, $f_p$, $\bar{f}_{p'}$, and $g_k$ represent the
quark, the anti-quark, and the gluon distribution functions (i.e., the
thermal equilibrium distributions plus fluctuations) with quantum
numbers $p$, $p'$, and $k$, respectively.  Then, the collision terms
are vanishing with $\feq(p)$, $\fbareq(p')$, and $\gequiv(k)$ due to
the detailed balance, and the first nonvanishing terms are linear in
terms of $\delta f_p$, $\delta\bar{f}_{p'}$, and $\delta g_k$
(or $\chi_p$, $\bar{\chi}_{p'}$, and $\tilde{\chi}_k$).  Simple
calculations lead us to
\begin{align}
  & \frac{C[f]}{\beta E_z} = \int_{k,p'} |\calM_{p\to p'+k}|^2
  (2\pi)^4\delta^{(4)}(p\!-\!k\!-\!p')
  \feq(p)[1-\feq(p')][1+\gequiv(k)]
  (\chi_p-\tilde{\chi}_k-\chi_{p'}) \notag\\
  &\qquad + \int_{k,p'} |\calM_{p+k\to p'}|^2
  (2\pi)^4\delta^{(4)}(p\!+\!k\!-\!p')
  \feq(p) \gequiv(k)[1-\feq(p')]
  (\chi_p+\tilde{\chi}_k-\chi_{p'}) \notag\\
  &\qquad + \int_{k,p'} |\calM_{p+p'\to k}|^2
  (2\pi)^4\delta^{(4)}(p\!+\!p'\!-\!k)
  \feq(p) \fbareq(p')[1+\gequiv(k)]
  (\chi_p+\bar{\chi}_{p'}-\tilde{\chi}_k)
\end{align}
for quarks and
\begin{align}
  & \frac{\bar{C}[f]}{\beta E_z} = \int_{k,p'}|\calM_{p\to p'+k}|^2
  (2\pi)^4\delta^{(4)}(p\!-\!k\!-\!p')
  \fbareq(p)[1-\fbareq(p')][1+\gequiv(k)]
  (\bar{\chi}_p-\tilde{\chi}_k-\bar{\chi}_{p'}) \notag\\
  &\qquad + \int_{k,p'}|\calM_{p+k\to p'}|^2
  (2\pi)^4\delta^{(4)}(p\!+\!k\!-\!p')
  \fbareq(p) \gequiv(k) [1-\fbareq(p')]
  (\bar{\chi}_p+\tilde{\chi}_k-\bar{\chi}_{p'}) \notag\\
  &\qquad + \int_{k,p'}|\calM_{p+p'\to k}|^2
  (2\pi)^4\delta^{(4)}(p\!+\!p'\!-\!k)
  \fbareq(p) \feq(p') [1+\gequiv(k)]
  (\bar{\chi}_p+\chi_{p'}-\tilde{\chi}_k)
\end{align}
for anti-quarks, and
\begin{align}
  & \frac{\tilde{C}[f]}{\beta E_z} = \int_{p,p'}|\calM_{k\to p+p'}|^2
  (2\pi)^4\delta^{(4)}(k\!-\!p\!-\!p')
  \gequiv(k)[1-\feq(p)][1-\fbareq(p')]
  (\tilde{\chi}_k-\chi_p-\bar{\chi}_{p'}) \notag\\
  &\qquad + \int_{p,p'}|\calM_{k+p\to p'}|^2
  (2\pi)^4\delta^{(4)}(k\!+\!p\!-\!p')
  \gequiv(k) \feq(p)  [1-\feq(p')]
  (\tilde{\chi}_k+\chi_p-\chi_{p'}) \notag\\
  &\qquad + \int_{p,p'}|\calM_{k+p'\to p}|^2
  (2\pi)^4\delta^{(4)}(k\!+\!p'\!-\!p)
  \gequiv(k)\fbareq(p')[1-\fbareq(p)]
  (\tilde{\chi}_k+\bar{\chi}_{p'}-\bar{\chi}_p)
\end{align}
for gluons.

Now our problem is reduced to the calculation of the scattering
amplitudes.   It is straightforward to write down the amplitude
contributions from the synchrotron radiation and the pair annihilation
processes as follows:
\begin{align}
  \ri \calM_{p\to k+p'} = \ri g\, \bar{u}(p')\gamma^\mu t_a u(p)
  \varepsilon^\ast_\mu(s,k)\,, \\
  \ri \calM_{p+p'\to k} = \ri g\, \bar{v}(p')\gamma^\mu t_a u(p)
  \varepsilon^\ast_\mu(s,k)\,.
\end{align}
We can make the sum of the squared amplitudes over the spin $s$, the
angular momentum $l$, and the color $c$ to find expressions in terms
of the Dirac structures~\eqref{eq:AB} of the propagator, i.e.,
\begin{align}
  \label{eq:rad}
  \sum_{s,l,c} |\calM_{p\to k+p'}|^2 &= -\delta_{ff'}  g^2\Nc C_F\,
  \tr\bigl[ \gamma_\mu S^{f}_n(p) \gamma^\mu S^{f}_{n'}(p-k) \bigr] \,,\\
  \label{eq:pair}
  \sum_{s,l,c} |\calM_{p+p'\to k}|^2 &= +\delta_{ff'}  g^2\Nc C_F\,
  \tr\bigl[ \gamma_\mu S^{f}_n(p) \gamma^\mu S^{f}_{n'}(p-k) \bigr] \,,
\end{align}
where the color factor is $C_F=(\Nc^2-1)/(2\Nc)$ and $p$ and $k$ in
the above expression refer to the momenta only.  
We replaced the polarization sum
$\sum_s \varepsilon^\ast_\mu(s,k)\varepsilon_\nu(s,k)$
 by $-\eta_{\mu\nu}$ thanks to the gauge invariance of the scattering amplitude,
and transformed the Dirac structures using
$\sum_{l,s} u(p)\bar{u}(p)=S^{f}_n(p)$ and
$\sum_{l,s}v(p)\bar{v}(p)=-S^{f}_n(-p)$.

One might think that the amplitudes for the synchrotron radiation and
the pair annihilation process may look identical from
Eqs.~\eqref{eq:rad} and \eqref{eq:pair} apart from the overall sign,
but the kinematic regions of involved momenta are quite different.
For the synchrotron radiation process of Eq.~\eqref{eq:rad} with
$p\to k+p'$, the on-shell condition reads,
$\varepsilon_{f n} - \omega_k - \varepsilon_{f n'}=0$
(which is possible for $n>n'$), from which we
can solve $\xi_k^{f} = |\bk_\perp|^2/(2|q_f B|)$, that is given as
\begin{equation}
  \xi^{f}_k = \xi^{f}_- = \frac{(\varepsilon_{f n}-\varepsilon_{f n'})^2
    - (p_z-p_z')^2}{2|q_{f}B|}\,.
\end{equation}
This quantity should be non-negative from the definition of
$\xi^{f}_k$, so that the integration range of $p_z$ and $p_z'$ should
be restricted to satisfy,
\begin{equation}
  m_{f n}^2 {p_z'}^2 + m_{f n'}^2 p_z^2 - p_z p_z' (m_{f n}^2 + m_{f n'}^2)
  - \frac{1}{4}(m_{f n}^2 - m_{f n'}^2)^2 \le 0\,.
\end{equation}
For given $p_z$, equivalently, we find that $p_z'$ ranges in
$p_{z-}' \le p_z' \le p_{z+}'$, where
\begin{equation}
  p_{z\pm}' = p_z \frac{m_{f n}^2 + m_{f n'}^2}{2m_{f n}^2}
  \pm \frac{m_{f n}^2-m_{f n'}^2}{2m_{f n}^2}\sqrt{m_{f n}^2+p_z^2}\,.
  \label{eq:pzpm}
\end{equation}
Thus, the phase space integration for the synchrotron radiation
process is rewritten as
\begin{equation}
\begin{split}
&  \int \frac{\rd^3 k}{(2\pi)^3} \frac{1}{2\omega_k}
  (2\pi)\delta(\omega_k-\varepsilon_{f n}+\varepsilon_{f n'}) \cdots\\
&  = \frac{1}{2}\vartheta(n-n'-1) \int\frac{\rd k_z}{2\pi}
  \int \rd \xi_k^{f}\, \delta(\xi_k^{f}-\xi_-^{f})\,
  \theta(p_{z+}'-p_z')\theta(p_z'-p_{z-}') \cdots\,,
  \end{split}
\end{equation}
where $\vartheta(x)$ is the unit step function,
\begin{equation}
  \vartheta(x) := \begin{cases}
  0 & (x<0) \\ 1 & (x\geq 0)
  \end{cases} \,.
\end{equation}
In the same way, we can identify the phase space for the pair
annihilation process of Eq.~\eqref{eq:pair} with $p+p'\to k$.  From
the energy on-shell condition, $\xi_k^{f}$ is fixed as
\begin{equation}
  \label{eq:xi+}
  \xi_k^{f} = \xi_+^{f} = \frac{(\varepsilon_{f n} + \varepsilon_{f n'})^2
    - (p_z + p_z')^2}{2|q_{f}B|}\,.
\end{equation}
It is easy to confirm that $\xi_+^{f}$ is positive definite, and so
the integrations of $p_z$ and $p_z'$ are not limited unlike the
synchrotron radiation.  Then, the phase space integration for the pair
annihilation process should be
\begin{equation}
  \int \frac{\rd^3 k}{(2\pi)^3} \frac{1}{2\omega_k}
  (2\pi)\delta(\omega_k-\varepsilon_{f n}-\varepsilon_{f n'}) \cdots
  = \frac{1}{2}\int\frac{\rd k_z}{2\pi}
  \int \rd \xi_k^{f}\, \delta(\xi_k^{f}-\xi^{f}_+) \cdots\,.
\end{equation}
These assignments of allowed kinematic regions lead us to the
following distinctions in the collision terms as
\begin{align}
  \label{eq:MX}
  & \int_{k,p,p'} \!\!\!\!\!|\calM_{p\to p'+k}|^2
  (2\pi)^4\delta^{(4)}(k\!-\!p\!+\!p')
  = -\frac{1}{2}\sum_{f,\,n>n'} \int\frac{\rd p_z}{2\pi}\frac{1}{2\varepsilon_{f n}}
  \int'\frac{\rd p_z'}{2\pi}\frac{1}{2\varepsilon_{f n'}} X(n,n',\xi_-), \\
  & \int_{k,p,p'} |\calM_{p+p'\to k}|^2 (2\pi)^4\delta^{(4)}(p\!+\!p'\!-\!k)
  = \frac{1}{2}\sum_{f,\,n,n'} \int\frac{\rd p_z}{2\pi} \frac{1}{2\varepsilon_{f n}}
  \int\frac{\rd p_z'}{2\pi} \frac{1}{2\varepsilon_{f n'}} X(n,n',\xi_+),
\end{align}
where the common integrand is given from Eqs.~\eqref{eq:rad} and
\eqref{eq:pair} as
\begin{equation}
  X(n,n',\xi_k) := g^2 \Nc C_F \int\frac{\rd^2 p_\perp}{(2\pi)^2}
  \tr\bigl[ \gamma_\mu S^{f}_n(p) \gamma^\mu S^{f}_{n'}(p-k) \bigr] \,.
\end{equation}
Here, in Eq.~\eqref{eq:MX}, we introduced a compact notation for the
momentum integration as
\begin{equation}
  \int' \frac{\rd p_z'}{2\pi} := \int_{p_{z-}'}^{p_{z+}'}
  \frac{\rd p_z'}{2\pi}
  \label{eq:primed}
\end{equation}
using $p_{z\pm}'$ defined in Eq.~\eqref{eq:pzpm}.  Because the final
form of $X(n,n',\xi)$ is complicated, let us first give the final form
here, and then look at major steps of the derivation.  The final
expression reads,
\begin{align}
  \label{eq:X}
  X(n,n',\xi) &= g^2\Nc C_F \frac{|q_{f}B|}{2\pi}\,
  \rme^{-\xi}\frac{n!}{n'!}\, \xi^{n'-n} \biggl\{ \biggl[ 4m_{f}^2
  - 4|q_{f}B|(n+n'-\xi)\frac{1}{\xi}(n+n')\biggr] F(n,n',\xi) \notag\\
  &\qquad\qquad\qquad\qquad + 16|q_{f}B|n'(n+n')\frac{1}{\xi}
    L_n^{(n'-n)}(\xi) L_{n-1}^{(n'-n)}(\xi)\biggr\} \,,
\end{align}
where we introduced a new function, $F(n,n',\xi)$, defined by
\begin{equation}
  F(n,n',\xi)  :=
  \begin{cases}
    \displaystyle 1 & (n=0) \\
    \displaystyle \bigl[ L_n^{(n'-n)}(\xi) \bigr]^2
    + \frac{n'}{n} \bigl[ L_{n-1}^{(n'-n)}(\xi) \bigr]^2
    & (n>0) \,.
  \end{cases}
\end{equation}
All calculational details would be quite lengthy, but we shall give a
sketch of how to arrive at the above final form from
Eqs.~\eqref{eq:rad} and \eqref{eq:pair} so that interested readers could
reproduce it.  Using Eq.~\eqref{eq:AB} a common building block of
Eqs.~\eqref{eq:rad} and \eqref{eq:pair} is expressed as
\begin{equation}
  \gamma_\mu S^{f}_n(p)\gamma^\mu
  = 2\bigl[ (-P^{f}_- \, \Slash{p}_\parallel + m_{f}) A_{n+}
  + (-P^{f}_+ \, \Slash{p}_\parallel + m_f) A_{n-}
  - \Slash{p}_\perp B_n  \bigr]\,.
\end{equation}
Taking the Dirac trace in Eqs.~\eqref{eq:rad} and \eqref{eq:pair} is
straightforward, which yields,
\begin{equation}
\label{eq:traceSn}
  \begin{split}
  & \tr\bigl[ \gamma_\mu S^{f}_n(p)\gamma^\mu S^{f}_{n'}(p-k)\bigr] \\
  &= 4m_{f}^2 \bigl[ A_{n'+}(4\xi_{p-k}^{f}) A_{n+}(4\xi_p^{f})
      + A_{n'-}(4\xi_{p-k}^{f}) A_{n-}(4\xi_p^{f}) \bigr] \\
  &\qquad - 4 \bigl[(p_\parallel-k_\parallel)\cdot p_\parallel-m_{f}^2\bigr]
    \bigl[ A_{n'-}(4\xi_{p-k}^{f}) A_{n+}(4\xi_p^{f})
      + A_{n'+}(4\xi_{p-k}^{f}) A_{n-}(4\xi_p^{f}) \bigr] \\
  &\qquad + 8(\bp_\perp-\bk_\perp)\cdot \bp_\perp
    B_{n'}(4\xi_{p-k}^{f}) B_n(4\xi_p^{f})\,.
  \end{split}
\end{equation}
We can slightly simplify the second line after the equality in the
above equation using the on-shell condition as
\begin{equation}
  2\bigl[(p_\parallel-k_\parallel)\cdot p_\parallel-m_{f}^2\bigr]
  = p_\parallel^2-m_{f}^2+(p_\parallel-k_\parallel)^2-m_{f}^2-k_\parallel^2
  = 2|q_{f}B| (n+n' - \xi_k^{f})\,,
\end{equation}
where we used $k_\parallel^2=\bk_\perp^2=2|q_{f}B|\xi_k^{f}$ (i.e.,
onshell-ness of massless gluons).  To perform the integration of
Eq.~\eqref{eq:traceSn} with respect to the transverse momentum, we
need formulas of (generalized) Laguerre polynomials:
\begin{align}
  \Ffunc_{n,n'}(\xi_{k}^{f}) &:= 2|q_{f}B|(-1)^{n-n'}
  \int \frac{\rd ^2 p_\perp}{(2\pi)^2}\, \rme^{-2\xi_p^{f}}L_n(4\xi_p^{f})\,
  \rme^{-2\xi_{p-k}^{f}}L_{n'}(4\xi_{p-k}^{f}) \notag\\
  &= \frac{|q_{f}B|}{2\pi} \rme^{-\xi_k^{f}}\frac{n!}{n'!}(\xi_k^{f})^{n'-n}
  [L_n^{(n'-n)}(\xi_k^{f})]^2 \,,\\
  \Gfunc_{n,n'}(\xi_k^{f}) &:= 2|q_{f}B|(-1)^{n-n'}
  \int \frac{\rd^2p_\perp}{(2\pi)^2}\, \rme^{-2\xi_p^{f}}L^{(1)}_{n-1}(4\xi_p^{f})\,
  \rme^{-2\xi_{p-k}^{f}}L^{(1)}_{n'-1}(4\xi_{p-k}^{f})
  \bp_\perp \cdot (\bp_\perp-\bk_\perp) \notag\\
  &= \frac{|q_{f}B|}{2\pi} \rme^{-\xi_k^{f}}\frac{n!}{n'!}(\xi_k^{f})^{n'-n}
  n' L_{n-1}^{(n'-n)}(\xi_k^{f})L_{n}^{(n'-n)}(\xi_k^{f}) \,.
\end{align}
With these formulas and the explicit forms of Eqs.~\eqref{eq:An+},
\eqref{eq:An-}, and \eqref{eq:Bn}, we find,
\begin{align}
  & \int\frac{\rd^2 p_\perp}{(2\pi)^2}\, \tr\bigl[
    \gamma_\mu S^{f}_n(p)\gamma^\mu S^{f}_{n'}(p-k) \bigr] \notag\\
  &= 4m_{f}^2 (\Ffunc_{n,n'}+\Ffunc_{n-1,n'-1})
    - 4|q_{f}B| (n+n' - \xi_k^{f}) (\Ffunc_{n-1,n'}
    +\Ffunc_{n,n'-1}) + 16|q_{f}B|\Gfunc_{n,n'} \notag\\
  &= \frac{|q_{f}B|}{2\pi}\,\rme^{-\xi_k^{f}}\frac{n!}{n'!}\,(\xi_k^{f})^{n'-n}
    \Biggl[4m_{f}^2 F(n,n',\xi_k^{f}) - 4|q_{f}B| (n+n' - \xi_k^{f})
      \frac{1}{\xi_k^{f}}\biggl\{ \frac{(\xi_k^{f})^2}{n}
      \bigl[L_{n-1}^{(n'-n+1)}(\xi_k^{f})\bigr]^2 \notag\\
  &\qquad\qquad\qquad\qquad
    + n' \bigl[L_n^{(n'-n-1)}(\xi_k^{f})\bigr]^2 \biggr\}
    + 16|q_{f}B| n'L_{n-1}^{(n'-n)}(\xi_k^{f}) L_n^{(n'-n)}(\xi_k^{f}) \Biggr]\,.
\end{align}
Using the following relation,
\begin{equation}
  \frac{\xi^2}{n}[L_{n-1}^{(n'-n+1)}(\xi)]^2
  + n' [L_n^{(n'-n-1)}(\xi)]^2
  = (n'+n)F(n,n',\xi)-4n'L_n^{(n'-n)}(\xi)L_{n-1}^{(n'-n)}(\xi)\,,
\end{equation}
we eventually obtain $X(n,n',\xi)$ in a form as given in Eq.~\eqref{eq:X}.

\subsection{Matrix elements}
\label{sec:evaluation}

We should solve $\chi$ to obtain the longitudinal conductivity from
Eq.~\eqref{eq:sigmazz}, and for this, we should invert the collision
operator $\calL$.  In actual numerical calculations we need to
represent the operator in a matrix form with appropriate bases.  Once
we prepare complete sets of $\eta$ and $\chi$, using the inner product
we defined in Eq.~\eqref{eq:inner}, we can express $(\eta,\calL\chi)$
in the language of the matrix algebra as
\begin{equation}
  \begin{split}
    (\eta,\calL\chi) = & 
     \sum_{n_{1},n_{2},l_{1},l_{2},f_{1},f_{2}}\Bigl(c^\eta_{f_{1}n_1 l_1}\calL^{qq}_{f_{1} n_1 l_1;f_{2} n_2 l_2}c^\chi_{f_{2} n_2 l_2}
    + c^\eta_{f_{1} n_1 l_1}\calL^{q\bar{q}}_{f_{1} n_1 l_1;f_{2} n_2 l_2}\bar{c}^\chi_{f_{2} n_2 l_2 } \\
    &\qquad\qquad\qquad +\bar{c}^\eta_{f_{1} n_1 l_1 }\calL^{\bar{q}q}_{f_{1} n_1 l_1;f_{2} n_2 l_2}c^\chi_{f_{2} n_2 l_2}
    +\bar{c}^\eta_{ f_{1} n_1 l_1 }\calL^{\bar{q}\bar{q}}_{f_{1} n_1 l_1;f_{2} n_2 l_2}\bar{c}^\chi_{f_{2} n_2 l_2}  
    \Bigr)\\[2ex]
    &+\sum_{n,m,l_{1},l_{2},f}\Bigl(\bar{c}^\eta_{f n l_1}\calL^{\bar{q}g}_{f n l_1;m l_2}\tilde{c}^\chi_{m l_2}  
    + \tilde{c}^\eta_{m l_1}\calL^{gq}_{m l_1;f n l_2}c^\chi_{f n l_2}\\
    &\qquad\qquad\quad+ \tilde{c}^\eta_{m l_1}\calL^{g\bar{q}}_{m l_1;f n l_2}\bar{c}^\chi_{f n l_2}  
      + c^\eta_{f n l_1}\calL^{qg}_{f n l_1;m l_2}\tilde{c}^\chi_{m l_2} \Bigr)\\[2ex]
  &+ \sum_{m_{1},m_{2},l_{1},l_{2}}\tilde{c}^\eta_{m_1 l_1}\calL^{gg}_{m_1 l_1;m_2 l_2}\tilde{c}^\chi_{m_2 l_2}  
    \Bigr]\,,
  \end{split}
\end{equation}
where $(c^\eta_{fnl},\bar{c}^\eta_{fnl},\tilde{c}^\eta_{ml})$ and
$(c^\chi_{fnl},\bar{c}^\chi_{fnl},\tilde{c}^\chi_{ml})$ represent the
components of $\eta=(\eta_p,\bar{\eta}_p,\tilde{\eta}_k)^T$ and
$\chi=(\chi_p,\bar{\chi}_p,\tilde{\eta}_k)^T$, respectively, with
chosen bases.  The simplest choice of the bases would be the
polynomial one.  These vectors are then expanded as
\begin{equation}
  \chi = 
  \begin{pmatrix}
    \displaystyle \sum_{f,n,l}c^\chi_{fnl}\; \delta^{nf} d_l(p_z) \\[3ex]
                    \displaystyle \sum_{f,n,l}\bar{c}^\chi_{fnl}\; \delta^{nf} d_l(p_z)\\[3ex]
                    \displaystyle \sum_{m,l}\tilde{c}^\chi_{ml}\;
                    d_l(k_z) b_{m}(k_{\perp})
    \end{pmatrix} \,,\qquad
  \eta = 
  \begin{pmatrix}
    \displaystyle \sum_{f,n,l}c^\eta_{fnl}\; \delta^{nf} d_l(p_z) \\[3ex]
                    \displaystyle \sum_{f,n,l}\bar{c}^\eta_{fnl}\; \delta^{nf} d_l(p_z)\\[3ex]
                    \displaystyle \sum_{m,l} \tilde{c}^\eta_{ml}\;
                    d_l(k_z) b_{m}(k_{\perp})
    \end{pmatrix}\,,
\end{equation}
where the $m$-th component of $\delta^{nf}$ is
$\delta^{n\times N_{f}+f}_m$ 
with the flavor number $N_{f}$, and
$d_l(p_z)$ and $b_m(k_\perp)$ are polynomial bases, namely,
$d_l(p_z) = \hat{p}_z |p_z|^l$ and $b_m(k_\perp) = k_{\perp}^{m}$.
Now, using these basis functions, we can explicitly write down the
matrix elements of the operator $\calL$.  The diagonal parts (i.e.,
$qq$, $\bar{q}\bar{q}$, and $gg$ components) are
\begin{align}
  & \calL^{qq}_{f_1 n_1 l_1;f_2 n_2 l_2 }
  =\delta_{f_{1} f_{2}}\Biggl\{
  -\frac{\delta_{n_1 n_2}}{2} \sum_{n'=0}^{n_1-1}
  \int\frac{\rd p_z}{2\pi} \frac{1}{2\varepsilon_{f_1 n_1}}
  \int'\frac{\rd p_z'}{2\pi} \frac{1}{2\varepsilon_{f_1 n'}}
    X(n_1,n',\xi_-)
    \feq(p)[1\!-\!\feq(p')] \notag\\
  &\quad\times [1\!+\!\gequiv(k)] d_{l_1}(p_z) d_{l_2}(p_z)
   - \frac{\delta_{n_1 n_2}}{2}\!\!\!
   \sum_{n=n_2+1}^\infty \int\frac{\rd p_z}{2\pi}
   \frac{1}{2\varepsilon_{f_1 n}} \int'\frac{\rd p'_z}{2\pi}
   \frac{1}{2\varepsilon_{f_1 n_2}}
    X(n,n_2,\xi_-)
    \feq(p) \notag\\
   &\quad\times [1\!-\!\feq(p')][1\!+\!\gequiv(k)]
    d_{l_1}(p'_z) d_{l_2}(p'_z)
    + \frac{\delta_{n_1 n_2}}{2}\sum_{n'=0}^\infty
    \int\frac{\rd p_z}{2\pi}\frac{1}{2\varepsilon_{f_1 n_1}}
    \int\frac{\rd p'_z}{2\pi}\frac{1}{2\varepsilon_{f_1 n'}}
     X(n_1,n',\xi_+) \notag\\
    &\quad\times  \feq(p)\fbareq(p')[1\!+\!\gequiv(k)] d_{l_1}(p_z) d_{l_2}(p_z)
    + \frac{\vartheta(n_1 \! - \! n_2 \!-\! 1)}{2}
    \int\frac{\rd p_z}{2\pi} \frac{1}{2\varepsilon_{f_1 n_1}}
      \int' \frac{\rd p'_z}{2\pi} \frac{1}{2\varepsilon_{f_1 n_2}}
      \notag\\
  &\quad\times X(n_1,n_2,\xi_-)
    \feq(p)[1\!-\!\feq(p')][1\!+\!\gequiv(k)] d_{l_1}(p_z) d_{l_2}(p'_z)
    + \frac{\vartheta(n_2 \!-\! n_1 \!-\! 1)}{2}
    \int\frac{\rd p_z}{2\pi} \frac{1}{2\varepsilon_{f_1 n_2}} \notag\\
   &\quad\times \int' \frac{\rd p'_z}{2\pi} \frac{1}{2\varepsilon_{f_1 n_1}}
    X(n_2,n_1,\xi_-)
    \feq(p)[1\!-\!\feq(p')][1\!+\!\gequiv(k)]d_{l_1}(p'_z) d_{l_2}(p_z) \Biggr\}\,,
\end{align}
\begin{align}
  & \calL^{\bar{q}\bar{q}}_{f_1 n_1 l_1;f_2 n_2 l_2}
  =\delta_{f_{1} f_{2}}\Biggl\{ -\frac{\delta_{n_1 n_2}}{2}\sum_{n'=0}^{n_1-1}
    \int\frac{\rd p_z}{2\pi} \frac{1}{2\varepsilon_{f_1 n_1}}
    \int'\frac{\rd p'_z}{2\pi} \frac{1}{2\varepsilon_{f_1 n'}}
    X(n_1,n',\xi_-)
    \fbareq(p)[1\!-\!\fbareq(p')] \notag\\
  &\quad\times [1\!+\!\gequiv(k)]d_{l_1}(p_z) d_{l_2}(p_z)
   - \frac{\delta_{n_1 n_2}}{2}\!\!\!\sum_{n=n_2+1}^\infty
    \int\frac{\rd p_z}{2\pi} \frac{1}{2\varepsilon_{f_1 n}}
    \int'\frac{\rd p'_z}{2\pi} \frac{1}{2\varepsilon_{f_1 n_2}}
    X(n,n_2,\xi_-)
    \fbareq(p) \notag\\
  &\quad\times [1\!-\!\fbareq(p')][1\!+\!\gequiv(k)]
    d_{l_1}(p'_z) d_{l_2}(p'_z)
    + \frac{\delta_{n_1 n_2}}{2}\sum_{n=0}^\infty
    \int\frac{\rd p_z}{2\pi} \frac{1}{2\varepsilon_{f_1 n}}
    \int\frac{\rd p'_z}{2\pi} \frac{1}{2\varepsilon_{f_1 n_2}}
    X(n,n_2,\xi_+) \notag\\
   &\quad\times \feq(p)\fbareq(p')[1\!+\!\gequiv(k)] d_{l_1}(p_z') d_{l_2}(p_z')
    + \frac{\vartheta(n_1 \!-\! n_2 \!-\!1)}{2}
     \int\frac{\rd p_z}{2\pi} \frac{1}{2\varepsilon_{f_1 n_1}}
     \int'\frac{\rd p'_z}{2\pi} \frac{1}{2\varepsilon_{f_1 n_2}}
     \notag\\
   &\quad\times X(n_1,n_2,\xi_-)
     \fbareq(p)[1\!-\!\fbareq(p')][1\!+\!\gequiv(k)] d_{l_1}(p_z) d_{l_2}(p'_z)
     + \frac{\vartheta(n_2 \!-\! n_1 \!-\!1)}{2}
     \int\frac{\rd p_z}{2\pi} \frac{1}{2\varepsilon_{f_1 n_2}}
     \notag\\
   &\quad\times \int'\frac{\rd p'_z}{2\pi} \frac{1}{2\varepsilon_{f_1 n_1}}
     X(n_2,n_1,\xi_-)
    \fbareq(p)[1\!-\!\fbareq(p')][1\!+\!\gequiv(k)] d_{l_1}(p'_z) d_{l_2}(p_z)\Biggr\}\,,
\end{align}
and
\begin{align}
  & \calL_{m_{1}l_{1};m_{2}l_{2}}^{gg}
  = \frac{1}{2}\sum_{f} \Biggl\{ -\sum_{n=1}^\infty \sum_{n'=0}^{n-1}
  \int\frac{\rd p_z}{2\pi} \frac{1}{2\varepsilon_{f n}}
  \int'\frac{\rd p'_z}{2\pi} \frac{1}{2\varepsilon_{f n'}}  
  X(n,n',\xi_-)
    \feq(p)[1-\feq(p')] \notag\\
  &\quad\times [1+\gequiv(k)]
    d_{l_1}(k_z) d_{l_2}(k_z) b_{m_1}(k_\perp)b_{m_2}(k_\perp)
  -\sum_{n=1}^\infty \sum_{n'=0}^{n-1}
  \int\frac{\rd p_z}{2\pi}\frac{1}{2\varepsilon_{f n}}
    \int\frac{\rd p'_z}{2\pi}\frac{1}{2\varepsilon_{f n'}} \notag\\
  &\quad\times X(n,n',\xi_-)
    \fbareq(p)[1-\fbareq(p')][1+\gequiv(k)]
    d_{l_1}(k_z) d_{l_2}(k_z) b_{m_1}(k_\perp)b_{m_2}(k_\perp) \notag\\
  &\quad + \sum_{n=0}^\infty \sum_{n'=0}^\infty
  \int\frac{\rd p_z}{2\pi}\frac{1}{2\varepsilon_{f n}}
  \int\frac{\rd p'_z}{2\pi}\frac{1}{2\varepsilon_{f n'}} X(n,n',\xi_+)
  \feq(p)\fbareq(p')[1+\gequiv(k)] \notag\\
  &\quad\times d_{l_1}(k_z) d_{l_2}(k_z) b_{m_1}(k_\perp)b_{m_2}(k_\perp)
    \Biggr\}\,.
\end{align}
In the same way the off-diagonal parts are
\begin{align}
  \calL^{q\bar{q}}_{f_{1} n_1 l_1;f_{2} n_2 l_2}
   & = \frac{\delta_{f_1f_2}}{2}\int\frac{\rd p_z}{2\pi}\frac{1}{2\varepsilon_{f_1 n_1}}
    \int\frac{\rd p'_z}{2\pi}\frac{1}{2\varepsilon_{f_{1} n_2}}
    \notag\\
  &\quad\times X(n_1,n_2,\xi_+)
    \feq(p)\fbareq(p')[1+\gequiv(k)]  d_{l_1}(p_z) d_{l_2}(p'_z)\,,
\end{align}
\begin{align}
  & \calL^{qg}_{f n_1 l_1;m l_2}
  = \frac{1}{2}\sum_{n=n_1+1}^\infty \int\frac{\rd p_z}{2\pi}
  \frac{1}{2\varepsilon_{f n}} \int\frac{\rd p'_z}{2\pi}
  \frac{-1}{2\varepsilon_{f n_1}} X(n,n_1,\xi_-)
  \feq(p)[1-\feq(p')][1+\gequiv(k)] \notag\\
  &\quad\times d_{l_1}(p'_z) d_{l_2}(k_z) b_{m}(k_\perp)
   -\frac{1}{2}\sum_{n'=0}^{n_1-1} \int\frac{\rd p_z}{2\pi}
  \frac{1}{2\varepsilon_{f n_1}} \int\frac{\rd p'_z}{2\pi}
  \frac{-1}{2\varepsilon_{f n'}} X(n_1,n',\xi_-)
    \feq(p)[1-\feq(p')] \notag\\
  &\quad\times [1+\gequiv(k)]
  d_{l_1}(p_z) d_{l_2}(k_z) b_{m}(k_\perp)
   -\frac{1}{2}\sum_{n'=0}^\infty \int\frac{\rd p_z}{2\pi}
  \frac{1}{2\varepsilon_{f n_1}} \int\frac{\rd p'_z}{2\pi}
  \frac{1}{2\varepsilon_{f n'}} X(n_1,n',\xi_+) \notag\\
  &\quad\times \feq(p)\fbareq(p')[1+\gequiv(k)]
  d_{l_1}(p_z) d_{l_2}(k_z) b_{m}(k_\perp) \,,
\end{align}
and
\begin{align}
  & \calL^{\bar{q}g}_{f n_1 l_1;m l_2}
  = \frac{1}{2}\sum_{n=n_1+1}^\infty \int\frac{\rd p_z}{2\pi}
  \frac{1}{2\varepsilon_{f n}} \int\frac{\rd p'_z}{2\pi}
  \frac{-1}{2\varepsilon_{f n_1}} X(n,n_1,\xi_-)
  \fbareq(p)[1-\fbareq(p')][1+\gequiv(k)] \notag\\
  &\quad\times d_{l_1}(p_z') d_{l_2}(k_z) b_{m}(k_\perp)
   -\frac{1}{2}\sum_{n'=0}^{n_1-1} \int\frac{\rd p_z}{2\pi}
  \frac{1}{2\varepsilon_{f n_1}} \int\frac{\rd p'_z}{2\pi}
  \frac{-1}{2\varepsilon_{f n'}} X(n_1,n',\xi_-)
    \fbareq(p)[1-\fbareq(p')] \notag\\
  &\quad\times [1+\gequiv(k)]
  d_{l_1}(p_z) d_{l_2}(k_z) b_{m}(k_\perp)
   -\frac{1}{2}\sum_{n=0}^\infty \int\frac{\rd p_z}{2\pi}
  \frac{1}{2\varepsilon_{f n}} \int\frac{\rd p'_z}{2\pi}
  \frac{1}{2\varepsilon_{f n_1}} X(n,n_1,\xi_+) \notag\\
  &\quad\times \feq(p) \fbareq(p')[1+\gequiv(k)]
  d_{l_1}(p_z') d_{l_2}(k_z) b_{m}(k_\perp) \,.
\end{align}
Other off-diagonal components are given by the symmetric properties of
$\calL$, that is,
\begin{equation}
     \calL^{\bar{q}q}_{f_{1}n_1l_1;f_{2}n_2l_2} = \calL^{q\bar{q}}_{f_{2}n_2l_2;f_{1}n_1l_1}\,,\quad
    \calL^{gq}_{n_1l_1;f_{2}n_2l_2} = \calL^{qg}_{f_{2}n_2l_2;n_1l_1}\,,\quad
    \calL^{g\bar{q}}_{n_1l_1;f_{2}n_2l_2} = \calL^{\bar{q}g}_{f_{2}n_2l_2;n_1l_1}\,.
 \end{equation}
Now that we get all the matrix elements, we can numerically take the
inverse to obtain $\calL^{-1}$, which should be multiplied to
$\calS$.  Then, we also need to know the matrix representation of
$\calS$ projected with the same chosen bases:
\begin{align}
  S_{fnl} &= \Nc \frac{|q_{f}B|}{2\pi} \alpha_n
  \int\frac{\rd p_z}{2\pi} \feq(p)[1-\feq(p)]
  \Bigl( q_{f}\frac{p_z }{\varepsilon_{f n}} - p_z\frac{n_e}{\calE+\calP_z} \Bigr)
  d_l(p_z) \,, \\
  \bar{S}_{fnl} &= \Nc \frac{|q_{f}B|}{2\pi} \alpha_n
  \int\frac{\rd p_z}{2\pi} \fbareq(p)[1-\fbareq(p)]
  \Bigl( -q_{f}\frac{p_z }{\varepsilon_{f n}} - p_z\frac{n_e}{\calE+\calP_z} \Bigr)
  d_l(p_z) \,, \\
  \tilde{S}_{ml} &= 2(\Nc^2-1) \int\frac{\rd^3 k}{(2\pi)^3}
  \gequiv(k)[1+\gequiv(k)] \Bigl( -k_z \frac{n_e}{\calE+\calP_z} \Bigr)
  d_l(k_z) b_m(k_\perp) \,. 
\end{align}
Using these expressions we can calculate
$\chi=\calQ\calL^{-1}\calQ\calS$, namely,
$(c_{fnl}^\chi, \bar{c}_{fnl}^\chi, \tilde{c}_{ml}^\chi)$ numerically,
and we reexpress $\sigma_\parallel=\beta(\calS,\chi)$ in
Eq.~\eqref{eq:sigmazz} with these components, which leads to
\begin{equation}
  \label{eq:sigma_long}
  \sigma_\parallel = 
  \beta\sum_{f,n,l}\bigl( c_{fnl}^\chi S_{fnl}
  + \bar{c}_{fnl}^\chi \bar{S}_{fnl}) +\beta\sum_{m,l} \tilde{c}_{ml}^\chi \tilde{S}_{ml}\,.
\end{equation}
This is our final expression for $\sigma_\parallel$ from the
analytical side, and we must now switch gear to numerical
computations.  Before proceeding to the numerical results, we would
point out a crucial difference between zero and finite densities.
For $\mu_{f}=0$, the solution satisfies
$c^{\chi}_{fnl}=-\bar{c}^{\chi}_{fnl}$, and $\tilde{c}^{\chi}_{ml}=0$
due to charge conjugation symmetry.  In this case, the conductivity of
the whole system can be simply a sum of contributions from different
flavor sectors and there appears no mixing term.  In contrast, for
$\mu_{f}\neq0$, different flavors mix together through the gluonic
matrix elements, $\calL^{qg}$ and $\calL^{\bar{q}g}$.

\subsection{Numerical results and approximation checks}
\label{sec:numerical}

We have spelled out all necessary formulas explicitly, and we will
present numerical results from now on.  In principle the Landau level sum goes
to infinity, but for numerical calculations we should truncate it at a
certain maximum value of $n$, which we denote $n_{\text{max}}$.  In
this work we check the convergence up to $n_{\text{max}}=6$ at zero
density, and for finite density we take $n_{\text{max}}=4$, for the
computation is much demanding at finite density.  Also, in practice, we should truncate
the sum in the angular momentum $l$, for which we choose
$l_{\text{max}}=2$.  We have numerically confirmed that this $l$ sum
has fast convergence and $l_{\text{max}}=2$ is a very good
approximation.  For all the numerical results presented below, we
fix the QCD coupling constant as $g^2/(4\pi)=0.3$, the system
temperature as $T=200\MeV$ which is well above the QCD phase
transition.

\begin{figure}
  \centering 
  \includegraphics[width=0.7\textwidth]{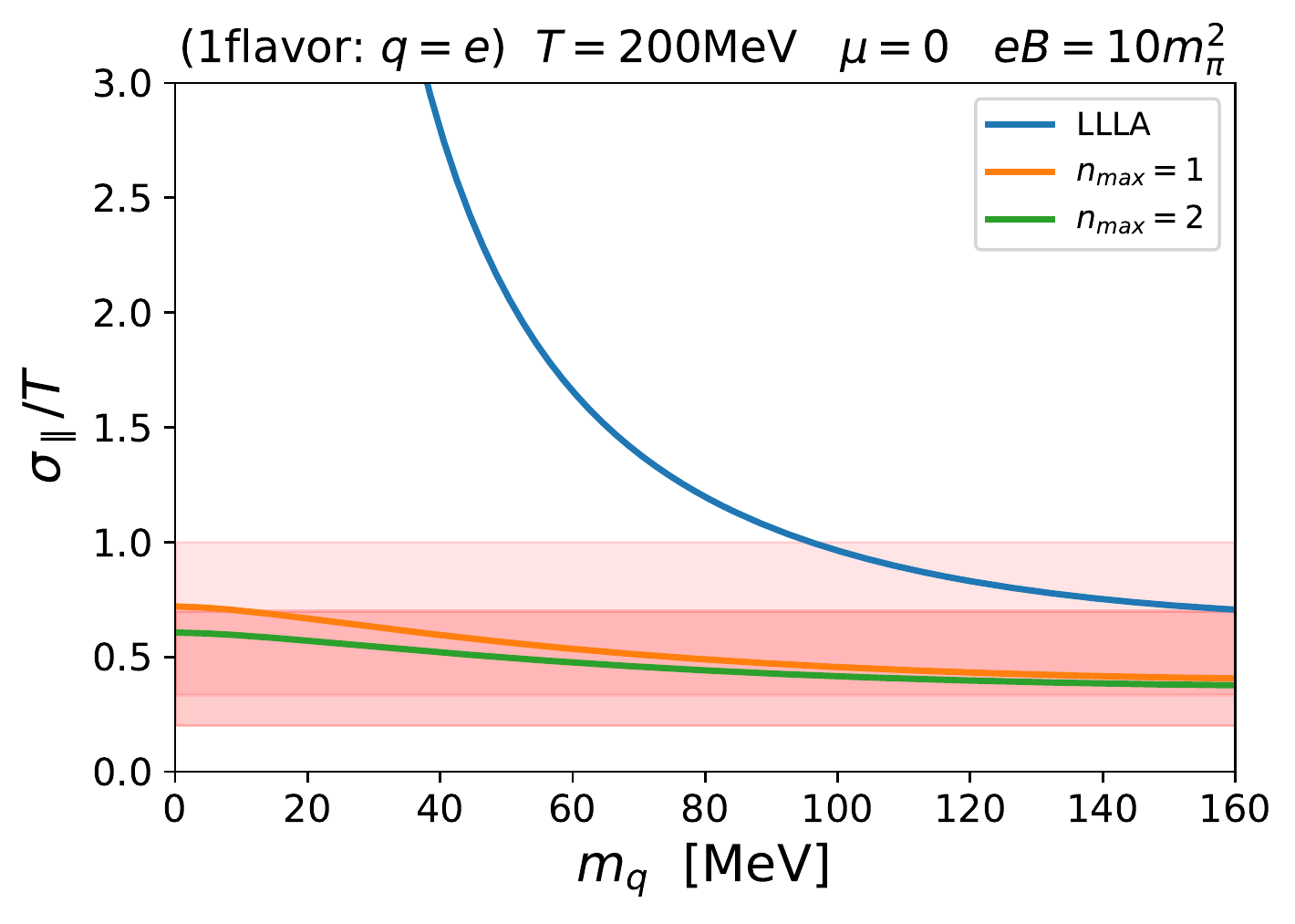}
  \caption{Mass dependence of the longitudinal electric conductivity
    for a single flavor case with $q_f=e$.  In the small mass region
    the LLLA blows up, while the results at $n_{\text{max}}=1,\,2$
    have regular behavior at $m_q\sim 0$.  The red shaded bands are the
    lattice-QCD estimates.  See the text for discussions.}
  \label{fig:massDep}
\end{figure}

Figure~\ref{fig:massDep} shows our numerical results for
$\sigma_\parallel/T$ as a function of the current quark mass, $m_q$,
for a single flavor case.  For convenience we set the electric charge
carried by this flavor as $q=e$.  The magnetic field is chosen to be
$eB=10m_\pi^2$ and the density is vanishing with $\mu=0$ for the
moment.  We will discuss the density dependence later.

It is evident from Fig.~\ref{fig:massDep} that the LLLA has rapid
enhancement as $m_q$ approaches zero.  In this way we can clearly
understand that the LLLA breaks down for $\sigma_\parallel(m_q\to 0)$
even though the magnetic field, $eB=10m_\pi^2$, is stronger than
typical QCD energy scales $\sim m_\pi$.  The singular behavior is, as
we explained in the introduction, attributed to the phase space
restriction.  In the LLLA the longitudinal and the transverse dynamics
of fermions are decoupled, and the longitudinal scattering in $(1+1)$
dimensions is prohibited for massless fermions due to the
energy-momentum conservation.  In this sense, only the LLLA is
exceptional, and the convergence of the Landau level sum is pretty
fast if one goes beyond the LLLA, as seen in Fig.~\ref{fig:massDep};
the $n_{\text{max}}=1$ results already give a good approximation close
to the $n_{\text{max}}=2$ results.  Here, we make one important
remark;  one might think that $\sigma_\parallel(m_q\to 0)$ should
diverge even beyond the LLLA because, according to the axial Ward
identity, the chirality in the massless limit is linearly increasing
with time regardless the scattering (except for the sphaleron
transition which is suppressed and negligible at weak coupling).  This
argument is mathematically correct, but physically the divergence
signifies a hydrodynamic mode (in a particular hydrodynamic regime;
see discussions in Introduction).  In fact, as we closely explained, the
conserved quantities such as the energy momentum tensor and the
electric charge constitute the hydrodynamic modes which should be
subtracted;  otherwise, they lead to divergence.  In the small mass
limit the axial charge is approximately conserved, thus it forms
another hydrodynamic mode.  We did not explicitly subtract this
additional hydrodynamic mode, but our calculation procedures without
coupling to the axial charge automatically drops out such a
hydrodynamic mode.  More specifically, out treatment of the
distributions did not allow for the axial charge beyond the linear
response regime, so that the hydrodynamic mode corresponding to the
axial charge is excluded.

In Fig.~\ref{fig:massDep} the red shaded bands represent the
lattice-QCD estimates.  It is interesting that our results are
quantitatively consistent with the lattice-QCD estimates, i.e.,
the light red region of $1/3 \le \sigma/T \le 1$ at
$T=1.45\,T_{\rm c}$ (for the quark charge squared sum
$C_{\text{em}}=1$ which corresponds to our single flavor with $q=e$)
in Ref.~\cite{Ding:2010ga} and the darker red region of
$0.2 \le \sigma/T \le 1$ at $T=1.1\,T_{\rm c}$ in
Ref.~\cite{Ding:2016hua}.  We note that $T_{\rm c}$ is the critical
temperature in quenched QCD, which is substantially larger than the
QCD (pseudo) critical temperature.  It is thus difficult to translate
the lattice temperature, and if we plug the physical QCD critical
temperature into $T_{\rm c}$, our choice of $T=200\MeV$ is almost
comparable to the lattice temperature.  As we see next, the
$B$ dependence of $\sigma_\parallel/T$ is moderate, at most $30\%$
changes up to $eB\sim 10m_\pi^2$, not order of magnitude differences.
Therefore, it would be sensible to make such a comparison of our
results at $eB=10m_\pi^2$ and the lattice-QCD results at $B=0$.
Still, we should emphasize that this is not an apple-to-apple
comparison, and we should not overstate the agreement beyond the
qualitative level.  Nevertheless, the quantitative agreement in
Fig.~\ref{fig:massDep} implies for $B=0$ that $\sigma_\parallel/T <1$
as reported in Refs.~\cite{Aarts:2007wj,Ding:2010ga} would be favored
and a former large value in Ref.~\cite{Gupta:2003zh} is unlikely.

\begin{figure}
  \centering 
  \includegraphics[width=0.7\textwidth]{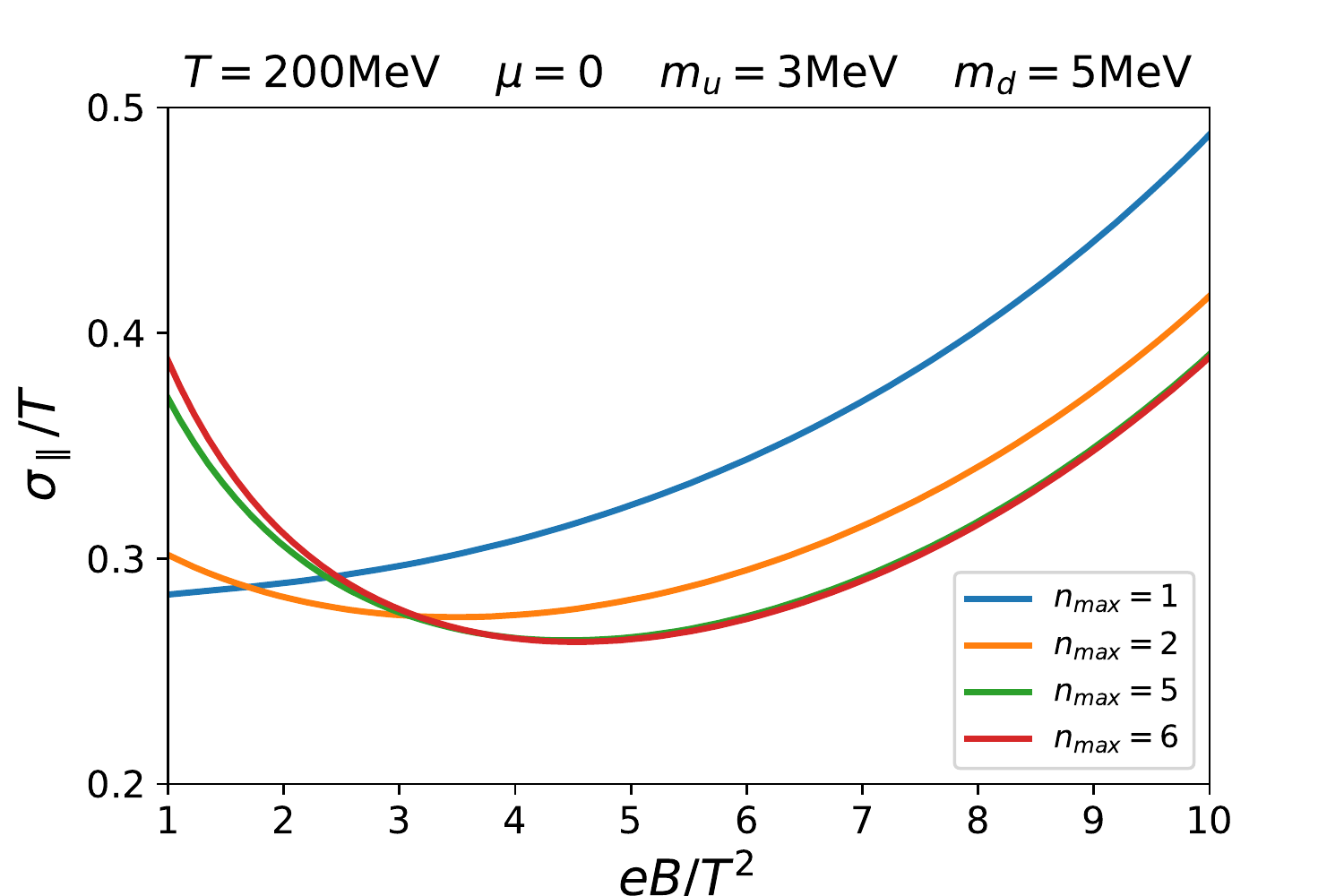}
  \caption{Magnetic dependence of the longitudinal electric
    conductivity for two favor case for various $n_{\text{max}}$.}
  \label{fig:BDep}
\end{figure}

Let us turn to the $eB$ dependence of $\sigma_\parallel/T$, which is
plotted in Fig.~\ref{fig:BDep}.  In this case we include $u$ and $d$ flavors
with physical quark masses ($m_u=3\MeV$ and $m_d=5\MeV$) and physical
electric charges ($q_u=\frac{2}{3}e$ and $q_d=-\frac{1}{3}e$).  If
$eB$ goes smaller, the convergence of the Landau level sum would
becomes slower, and in fact, Fig.~\ref{fig:BDep} shows that the
$n_{\text{max}}=1$ and the $n_{\text{max}}=2$ results behave
quite differently as $eB\to 0$.  Still, we can observe a tendency of
fast convergence from the $n_{\text{max}}=5$ and the
$n_{\text{max}}=6$ results for any magnetic strength.
Figure~\ref{fig:BDep} is our main result, and to emphasize its nature
as realization of the negative magnetoresistance, the resistance,
$\rho_\parallel=1/\sigma_{\parallel}$ was presented in
Fig.~\ref{fig:BDepLarge_rho} as displayed in the very beginning of
this paper.

It is a very interesting question how such a nonmonotonic shape
emerges;  actually the experimental data also shows a dip at small
magnetic field, which qualitatively agrees with our results.  In our
calculation the nonmonotonicy appears as a result of competing two
effects.  For large $eB$ the LLL contribution will become dominant, as
we closely check in the next subsection, and then $\sigma_\parallel$
is linearly proportional to $eB$, which can be intuitively understood
from the fact that the charge carrier increases then.  For small $eB$,
on the other hand, contributions from higher Landau levels lead to a
larger interaction cross section due to the phase space factor, which
lowers $\sigma_\parallel$ with increasing $eB$.  Interestingly, in the
intermediate region of $eB$, the behavior of $\sigma_\parallel$ looks
quadratic, as perceived from Fig.~\ref{fig:BDep}.  In this way our
results capture all qualitative features of the experimental data,
while quantitative details would depend on underlying theory.

\begin{figure}
  \centering 
  \includegraphics[width=0.7\textwidth]{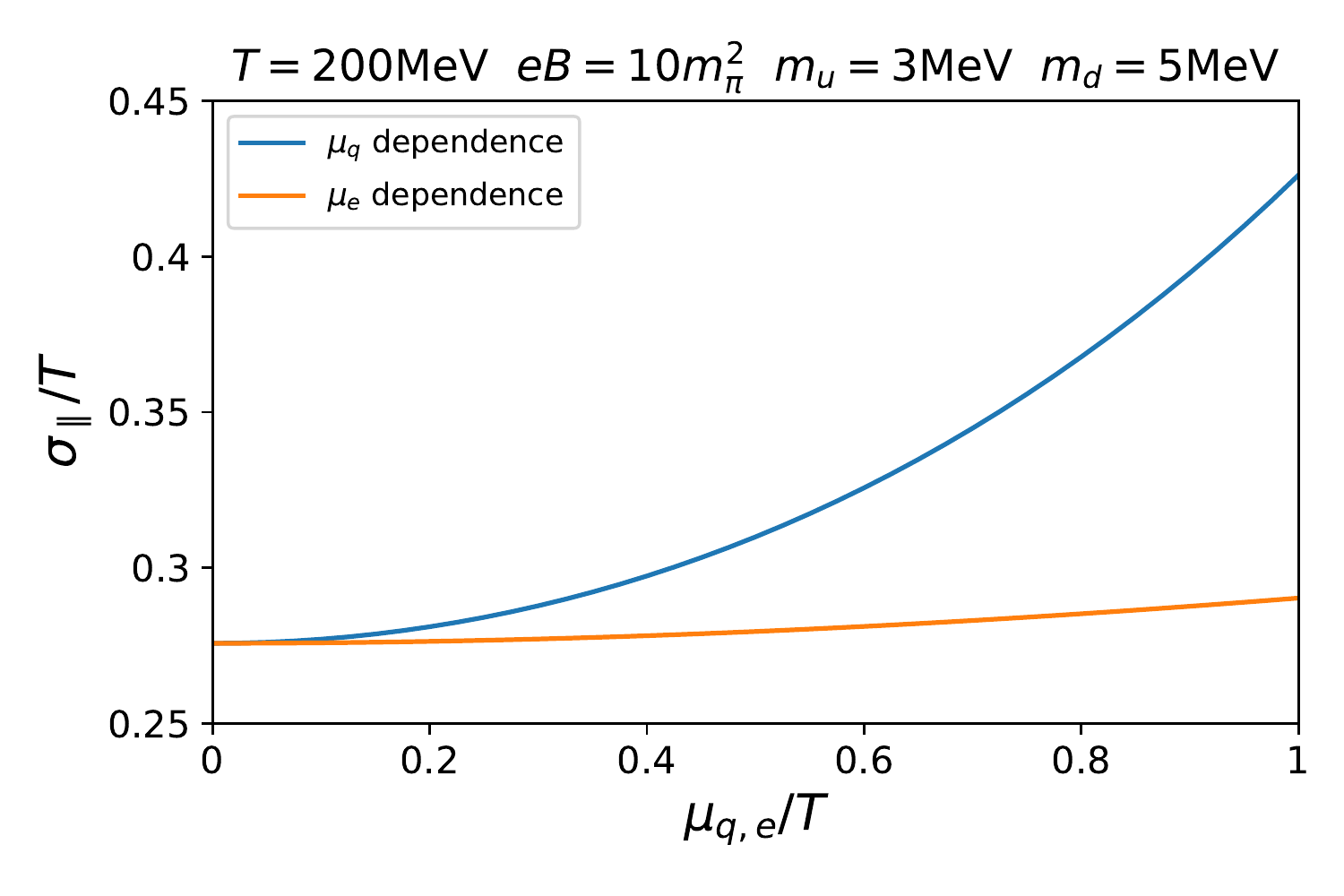}
  \caption{Chemical potential dependence of the longitudinal electric
    conductivity for two flavor case.  The upper line is for the quark
    chemical potential, while the lower one is for the electric chemical
    potential coupled to the electric charge.}
  \label{fig:muDep}
\end{figure}

Finally in this subsection, we shall consider the dependence of quark
chemical potential as shown in Fig.~\ref{fig:muDep}.  We can notice
that the dependence of the quark chemical potential $\mu_q$ is mild,
and that of the electric chemical potential $\mu_e$ is even more
suppressed.

\subsection{Asymptotics at strong magnetic field toward the LLLA}
\label{sec:recovery}

It would be an instructive check to confirm that the known results in 
the LLLA are correctly recovered from Eq.~\eqref{eq:sigma_long} in the 
limit of $q_{f}B\gg T^2$ (and $\mu_{f}=0$ for simplicity).  Under
this limit only the LLL contributes to physical
observables.  Since the synchrotron radiation requires changes with
respect to the
Landau levels, we can safely discard the radiation terms and the
absorption terms.  For the pair annihilation process,
$X(n=0,n'=0,\xi)$ in Eq.~\eqref{eq:X} simplifies as
\begin{equation}
  X(0,0,\xi_+^{f}) = 4m_f^2 g^2\Nc C_F \frac{|q_f B|}{2\pi} \rme^{-\xi_+^0}
\end{equation}
with
$\xi_+^0=[(\sqrt{p_z^2+m_f^2}+\sqrt{p_z'^2+m_f^2})^2-(p_z+p_z')^2]/(2|q_f B|)$
which is nothing but $\xi_+$ obtained from Eq.~\eqref{eq:xi+} with
$n=n'=0$.  When $|q_f B|$ is much larger than any other scales, we can
approximate $\rme^{-\xi_+^0} \approx 1$ neglecting the dependence on
$\xi_+^0$ which is suppressed by $|q_{f}B|$.  Then, the linearized
Boltzmann equations reduce to a simple form:
\begin{equation}
\begin{split}
  & q_f \Nc \frac{|q_f B|}{2\pi}\beta \feq(p)[1-\feq(p)]
  \frac{p_z}{\sqrt{p_z^2+m_f^2}} \\
  & = 4m_f^2 g^2\Nc C_F \beta \frac{|q_f B|}{2\pi} \cdot\frac{1}{2}
  \cdot\frac{1}{2\varepsilon_{f 0}}
  \int\frac{\rd p'_z}{2\pi}\frac{1}{2\varepsilon_{f 0}'}
  \feq(p) \fbareq(p')[1+\gequiv(k)] \chi_p\,, 
  \end{split}
\end{equation}
where $\varepsilon_{f 0}=\sqrt{p_z^2+m_f^2}$ and 
$\varepsilon_{f 0}'=\sqrt{p_z'^2+m_f^2}$.  Here, we do not have to
consider mixing terms with $\bar{\chi}_{p'}$ because $\bar{\chi}_{p'}$
is an odd function in terms of $p'$ and the integrand is even, so that
the integral vanishes.  Therefore, in this special limit, $\calL$ is
not really a matrix and we do not need to take its matrix inversion.
Actually, we can easily solve the above Boltzmann equation to find
$\chi_p$ as
\begin{equation}
  \chi_p = \frac{q_{f}}{g^2 C_F m_{f}^2} [1-\feq(p)]
  \frac{p_z}{\displaystyle \int\frac{\rd p'_z}{2\pi}
    \frac{1}{2\varepsilon_{f 0}'} \fbareq(p')[1+\gequiv(k)]} \,. 
\end{equation}
Thanks to the charge conjugation symmetry, the solution for the 
antiparticle is $\bar{\chi}_{p}=-\chi_{p}$.  Assembling these
expressions, we finally arrive at the longitudinal conductivity in the
LLLA, i.e.,
\begin{equation}
  \begin{split}
    \sigma_\parallel &= \sum_f\frac{\Nc \beta}{g^2 C_F m_{f}^2} 
    q_f^2\frac{|q_f B|}{2\pi}
    \int\frac{\rd p_z}{2\pi}\frac{p_z^2}{\sqrt{p_z^2+m_{f}^2}} 
    \frac{\feq(p)[1-\feq(p)]^2}
    {\displaystyle \int\frac{\rd p'_z}{2\pi}
     \frac{1}{\varepsilon_{f 0}'}\fbareq(p')[1+\gequiv(k)]} \,, 
 \end{split}
 \label{eq:LLLA}
\end{equation}
which is consistent with the results in Ref.~\cite{Hattori:2016lqx}.

\begin{figure}
  \centering 
  \includegraphics[width=0.7\textwidth]{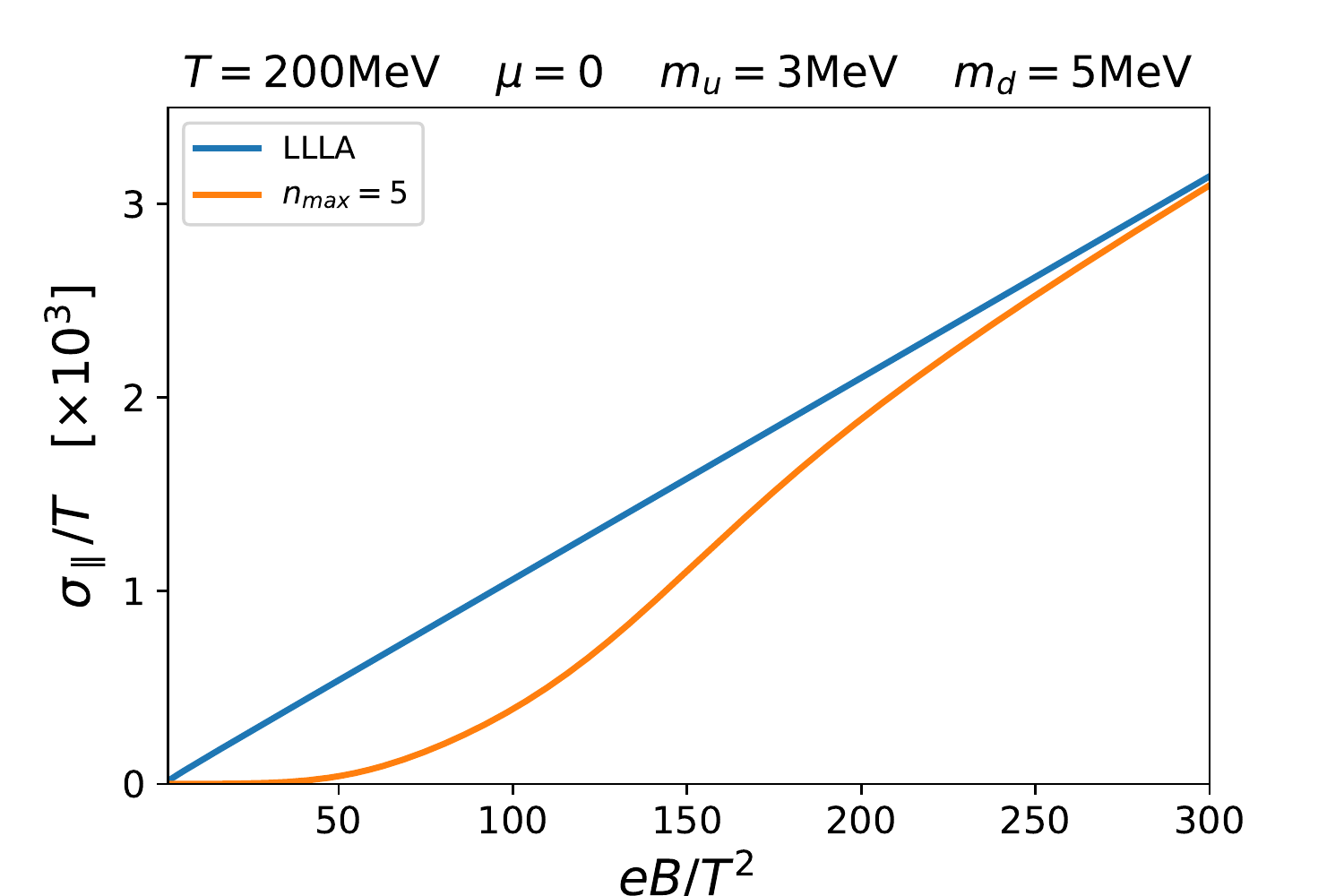}
  \caption{Magnetic field dependence of the longitudinal electric
    conductivity up to an asymptotically strong field region.  The
    full results (orange color) slowly approaches the results from the
    LLLA (blue color).}
  \label{fig:BDepLarge}
\end{figure}

Figure~\ref{fig:BDepLarge} is a numerical comparison between the LLLA
results (blue curve) and the full results (orange curve).  The latter
asymptotically approaches the former, but the convergence is very slow
(which quantitatively depends on the fermion masses).  For the present
parameters the LLLA can be a reasonable approximation only for
$eB \gtrsim 200T^2$.

We point out that Eq.~\eqref{eq:LLLA} is a quite suggestive expression
on the analytical level.  Obviously, $\sigma_\parallel$ diverges in
the limit of $m_f\to 0$, and this singular behavior implies that the
electric current also diverges.  In fact, in (1+1) dimensions, the
chiral anomaly predicts a topological current proportional to time for
non-interacting fermions, that is, $\partial_t j_z = \text{(const.)}$ or
$j_z \propto t$~\cite{Fukushima:2010vw,Copinger:2018ftr}.  In such a
case of linear rising current with time, the electric conductivity
defined at $k_0\to 0$ diverges.  Therefore, this divergent component
from the LLL should be attributed to the chiral anomaly.  One might
think that $\sigma_\parallel\to\infty$ is simply a consequence from
scatteringless nature of dimensionally reduced massless fermions, and
this kinematical argument is certainly correct.  Indeed, such a
picture of dimensionally reduced massless fermions provides us with a
classical description of the origin of the chiral
anomaly~\cite{Nielsen:1983rb}.  It is essentially important to
emphasize that the quantum anomaly is encoded in the Dirac equation in
the presence of gauge fields, so that we can in principle implement
contributions from the chiral anomaly in our current calculations.
Then, the diverging current induced by the topological effect from the
LLL is scattered off with higher Landau levels at finite mass
(corresponding to the effect of the relaxation time in
Refs.~\cite{Son:2012bg,Li:2014bha}) leading to a finite value of
$\sigma_\parallel$.

\section{Diagrammatic approach}
\label{sec:diagram}

Our calculation procedures in Sec.~\ref{sec:longitudinal} might have
looked quite different from direct application of the Kubo formula in
Sec.~\ref{sec:transverse}.  There, in the beginning of
Sec.~\ref{sec:longitudinal}, we gave handwaving arguments to relate
the Boltzmann equations and the Kubo formula through resummation, and
the purpose of this section is spelling out all the algebras to
establish this relation in a way as explicit as possible.

In the diagrammatic approach we need to solve the Bethe-Salpeter
equations and perform resummation for the propagator and the vertices
as illustrated in Fig.~\ref{fig:BSequation}.  We here limit our
consideration to the zero density ($\mu=0$) and the single flavor
($q=e$) case for notational brevity (dropping the flavor index
$f$ throughout this section).  If necessary, the generalization for
nonzero density and/or multiple flavors would be not so difficult.
Since we should deal with the
pinch singularities caused by the product of retarded and advanced
propagators, it would be convenient to employ  the ``$R/A$ basis'' in
the real-time
formalism~\cite{Aurenche:1991hi,vanEijck:1994rw,Hidaka:2010gh}.  With
the $R/A$ basis we can easily identify the pinch singularities out of the
diagrams.

\subsection{$R/A$ basis}

We can express the quark propagators, $S_{ab}(p)$ in the
Schwinger-Keldysh $1/2$ basis, appearing in Eq.~\eqref{eq:SRRK} and
nearby equations, in a matrix product form.  We can also represent the
gluon propagators, $G_{ab;\mu\nu}$ in the same way, and they
read~\cite{Hidaka:2010gh}:
\begin{equation}
\begin{split}
 S_{ab} (p) &= U_{a\alpha}(p_{0}) U_{b\beta}(-p_{0}) S_{\alpha\beta}(p),\\
 G_{ab;\mu\nu}(k) &= V_{a\alpha}(k_{0}) V_{b\beta}(-k_{0}) G_{\alpha\beta;\mu\nu}(k),
\end{split}
\end{equation}
with
\begin{equation}
S_{\alpha\beta}(p) =\begin{pmatrix}
0& -\ri S_{R}(p)\\
-\ri S_{A}(p) &0
\end{pmatrix},\qquad
G_{\alpha\beta;\mu\nu}=\begin{pmatrix}
0 & -\ri G_{R\mu\nu}(k)\\
-\ri G_{A\mu\nu}(k) &0
\end{pmatrix}.
\end{equation}
Here, the Latin ($a,b=1,2$) and Greek ($\alpha,\beta=R,A$) indices
represent the components in the $1/2$ and the $R/A$ bases,
respectively.  The transformation matrices are
\begin{equation}
U_{a\alpha}(p_{0}) =\begin{pmatrix}
 -n_{F}(p_{0}) & ~~ -\rme^{-\beta p_{0}}\\
 -n_{F}(p_{0}) &1
\end{pmatrix},\qquad
V_{a\alpha}(k_{0}) =\begin{pmatrix}
 n_{B}(k_{0}) & ~~ \rme^{-\beta k_{0}}\\
 n_{B}(k_{0}) &1
\end{pmatrix}.
\end{equation}
Now, the propagators are written in terms of the retarded and the
advanced ones, so that we can easily identify the pinch singularities
in the diagram.  Although the propagator matrix takes such a simple
form, the vertices are a little bit complicated.  We list up the
Feynman rules relevant for our present calculation below.  The quark
(straight line) and the gluon (curly line) propagators are as simple as
\begin{equation}
  \parbox[c]{2.cm}{\includegraphics[width=0.12\textwidth]{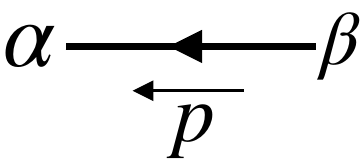}}
   = \, S_{\alpha\beta}(p)\,,\qquad
  \parbox[c]{2.cm}{\includegraphics[width=0.12\textwidth]{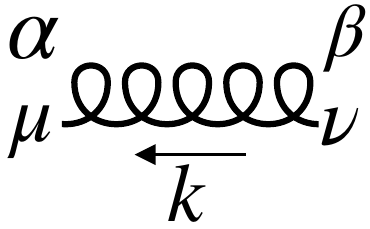}}
   = \, G_{\alpha\beta;\mu\nu}(k)\,.
\end{equation}
The quark-gluon (including the coupling constant $g$) and the
quark-photon (excluding the charge $q$) vertices involve the
Fermi-Dirac distribution function $n_F$ and the Bose-Einstein
distribution function $n_B$ for the $R$ components as
\begin{align}
\parbox[c]{2.4cm}{\includegraphics[width=0.15\textwidth]{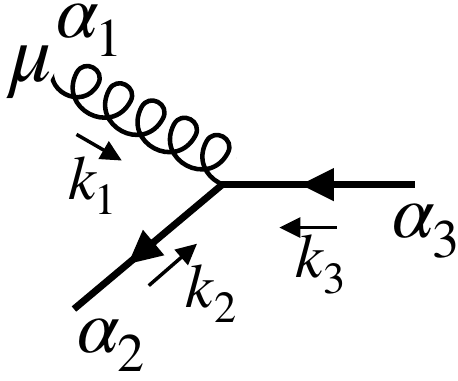}}
& =\ri g_{\alpha_{1}\alpha_{2}\alpha_{3}}(k_{1},k_{2},k_{3})\gamma^{\mu}t^{a} \notag\\
& = \ri g \gamma^{\mu} t^a \Bigl\{
  \bigl[1+n_{B}(k_{10})\bigr]^{\delta_{\alpha_{1},R}}
  \bigl[1-n_{F}(k_{20})\bigr]^{\delta_{\alpha_{2},R}}
  \bigl[1-n_{F}(k_{30})\bigr]^{\delta_{\alpha_{3},R}} \notag\\
  & \qquad\qquad -\bigl[n_{B}(k_{10})\bigr]^{\delta_{\alpha_{1},R}}
    \bigl[-n_{F}(k_{20})\bigr]^{\delta_{\alpha_{2},R}}
    \bigl[-n_{F}(k_{30})\bigr]^{\delta_{\alpha_{3},R}} \Bigr\}\,,
    \label{eq:vertexGluon}\\
\parbox[c]{2.4cm}{\includegraphics[width=0.15\textwidth]{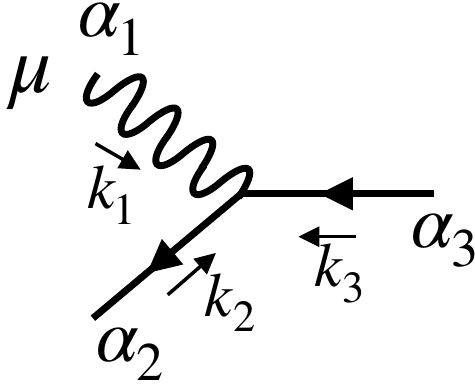}}
& = \Gamma_{\alpha_{1}\alpha_{2}\alpha_{3}}^{(0)\mu}(k_{1},k_{2},k_{3}) \notag\\
& = \gamma^{\mu}\Bigl\{
  \bigl[1+n_{B}(k_{10})\bigr]^{\delta_{\alpha_{1},R}}
  \bigl[1-n_{F}(k_{20})\bigr]^{\delta_{\alpha_{2},R}}
  \bigl[1-n_{F}(k_{30})\bigr]^{\delta_{\alpha_{3},R}} \notag\\
  & \qquad\qquad -\bigl[n_{B}(k_{10})\bigr]^{\delta_{\alpha_{1},R}}
    \bigl[-n_{F}(k_{20})\bigr]^{\delta_{\alpha_{2},R}}
    \bigl[-n_{F}(k_{30})\bigr]^{\delta_{\alpha_{3},R}} \Bigr\}\,. 
    \label{eq:vertexPhoton}
\end{align}
Here, $k_{10}, k_{20}, k_{30}$ denote the zeroth components (i.e.,
energies) of $k_1, k_2, k_3$, respectively.

\subsection{Expansion of the dressed propagator}
\label{sec:dressedPropagator}

The dressed propagator $S_{\alpha\beta}$ with insertion of self-energy
$\Sigma_{\alpha\beta}$ satisfies the following recursive equation:
\begin{equation}
  S_{\alpha\beta}  =  S^{(0)}_{\alpha\beta}
  + S^{(0)}_{\alpha\alpha'} \Sigma_{\alpha'\beta'} S_{\beta'\beta}\,,
\end{equation}
where $S^{(0)}_{\alpha\beta}$ represents the free propagator in the
presence of magnetic field.  For the retarded propagator, 
$S_R = \ri S_{RA}$, the above equation reduces to
\begin{equation}
 S_{R}  =    S^{(0)}_{R} -  S^{(0)}_{R}\Sigma_{R}S_{R}
\end{equation}
with 
$\Sigma_R := \ri \Sigma_{AR}$.  We can equivalently rewrite this into
a more familiar form as
\begin{equation}
  ( S^{(0)-1}_{R}+\Sigma_{R} ) S_{R}   = 1\,.
\label{eq:self-energy}
\end{equation}
Similarly, the dressed advanced  propagator should satisfy,
$S_{A} =  S^{(0)}_{A} -  S^{(0)}_{A}\Sigma_{A}S_{A}$.

We are going to solve Eq.~\eqref{eq:self-energy} perturbatively
treating $\Sigma_{R}$ as small perturbation.  At the zeroth order of
the expansion in terms of $\Sigma_R$, Eq.~\eqref{eq:self-energy} is
simply $S^{(0)-1}_{R} S_{R} = 1$.  For convenience of algebraic
transformations, let us introduce the complete basis as defined with the
Hamiltonian, $\gamma^{0} S^{(0)-1} = -\ri \partial_{0} + H$.  The
eigenvalue equation reads:
\begin{equation}
\gamma^{0} S^{(0)-1} |\qmn\rangle = \lambda_{\qmn} |\qmn\rangle\,,
\end{equation}
where $\qmn$ collectively represents a set of all quantum numbers,
i.e., $\qmn=(p_{0}, p_{z},\ph, n, \qm, s)$, and the eigenvalue is
$\lambda_{\qmn} = -p_{0}+ \ph \varepsilon_{n}$ with
$\varepsilon_{n}=\sqrt{p_{z}^{2}+2eBn+m^{2}}$.  In the present notation
$\ph$ is the quantum number for particle ($\ph=+1$) and antiparticle
($\ph=-1$), $s$ is the spin, and $\qm$ represents other degrees of
freedom characterizing the eigenstates which depend on the gauge
choice.  In the symmetric gauge $\qm$ amounts to the angular momentum
$l$ and the color $c$ as we considered for numerical calculations.  We
normalize the eigenstates as
\begin{equation}
  \langle \qmn|\qmn'\rangle = \delta_{\qmn,\qmn'}
  =  (2\pi)^{2}\delta^{(2)}(p_{\parallel}-p'_{\parallel})
  \delta_{n,n'}\delta_{\qm,\qm'}\delta_{s,s'}\delta_{\ph,\ph'} \,.
\end{equation}
The configuration space representation of states is
\begin{equation}
  \langle x| \qmn\rangle = \rme^{-\ri p_{\parallel}\cdot x_{\parallel}}
  \biggl[ \frac{\delta_{\ph,+}}{\sqrt{2\varepsilon_{n}}}
  u(p_{z}, {n}, \qm, s, \bx_{\perp}) +
  \frac{\delta_{\ph,-}}{\sqrt{2\varepsilon_{n}}}
  v(-p_{z}, {n}, \qm, s, \bx_{\perp}) \biggr] \,.
\end{equation}
Here, as usual, $u(p_{z},{n},\qm,s,\bx_{\perp})$ and
$v(-p_{z},{n},\qm,s,\bx_{\perp}) $ are wave functions of positive and
negative energy states solved from
\begin{equation}
\begin{split}
  H \rme^{\ri p_{z}z} u(p_{z},{n},\qm,\bx_{\perp})
  &= \varepsilon_{n}\, \rme^{\ri p_{z}z} u(p_{z},{n},\qm,\bx_{\perp})\,, \\
  H \rme^{\ri p_{z}z}v(-p_{z},{n},\qm,\bx_{\perp})
  &= -\varepsilon_{n}\, \rme^{\ri p_{z}z}v(-p_{z},{n},\qm,\bx_{\perp})\,.
\end{split}
\end{equation}

%
%

Using these notations we can construct the retarded propagator as
\begin{equation}
  S_{R}^{(0)}(x,x') = \sum_{\qmn}\langle x|\qmn\rangle
  \frac{1}{\lambda_{\qmn}-\ri \epsilon}  \overline{\langle\qmn|}x'\rangle\,,
\end{equation}
where we introduced
$\overline{\langle \qmn |} x'\rangle := \langle \qmn| x'\rangle \gamma^{0}$.
Generally $S_R^{(0)}(x,x')$ is not a function of $x-x'$ though the
magnetic field does not break translational invariance.  It is
well-known that, summing over $\qm$ and $s$, we can factorize
$S_R^{(0)}(x,x')$ as
\begin{equation}
  S_{R}^{(0)}(x,x') = e^{\ri\Theta(x,x')} \tilde{S}_{R}^{(0)}(x-x')
\end{equation}
into $\tilde{S}_R^{(0)}(x-x')$ [whose Fourier transform is found in
Eq.~\eqref{eq:SRA}] and $\Theta(x,x')$ is called the Schwinger
phase.  This phase plays no role, which is the case for our present 
problem, unless we consider operators
involving derivative coupling.
Therefore, we do not have to take care of the phase that
cancels out in the end, and drop the tilde from the propagator for
notational brevity.  For this translationally invariant part of the
propagator we can move to Fourier space in which the propagator takes
a form of
\begin{equation}
  \langle p|S_{R}^{(0)}|p' \rangle=S_{R}^{(0)}(p,p')
  = S_{R}^{(0)}(p)(2\pi)^{4}\delta^{(4)}(p-p')
  =  \sum_{\qmn}\langle p|\qmn\rangle
  \frac{1}{\lambda_{\qmn}-\ri \epsilon}\overline{\langle \qmn|}p'\rangle\,,
\end{equation}
where
\begin{equation}
 \begin{split}
  \langle p'|\qmn\rangle &=  \int \rd^{4}x\, \rme^{\ri p'\cdot x}
  \langle x|\qmn\rangle \\
   &= (2\pi)^{2}\delta^{(2)}(p_{\parallel}-p'_{\parallel})
\sqrt{V_{\perp}}  \biggl[ \frac{\delta_{t,+}}{\sqrt{2\varepsilon_{n}}} u(\bp',n,\qm,s)
  + \frac{\delta_{t,-}}{\sqrt{2\varepsilon_{n}}} v(-\bp',n,\qm,s) \biggr]
  \end{split}
\label{eq:pn}
\end{equation}
with Fourier transformed wave functions given as
\begin{equation}
  \begin{split}
    u(\bp,n,\qm) &:= \frac{1}{\sqrt{V_{\perp}}}\int \rd^{2}x_\perp \,
    \rme^{-\ri \bx_{\perp}\cdot\bp_{\perp}} u(p_{z},n,\ell,\bx_{\perp}) \,,\\
    v(-\bp,n,\qm) &:= \frac{1}{\sqrt{V_{\perp}}}\int \rd^{2}x_\perp \,
    \rme^{-\ri \bx_{\perp}\cdot\bp_{\perp}} v(-p_{z},n,\ell,\bx_{\perp}) \,.
    \label{eq:waveFunctions}
  \end{split}
\end{equation}
Here, $V_{\perp}$ is the transverse area on a plane perpendicular to the magnetic field,
and $1/\sqrt{V_\perp}$ is the normalization such that the Dirac structures satisfy 
$\sum_{l,s} u(p)\bar{u}(p)=S^{f}_n(p)$ and $\sum_{l,s}v(p)\bar{v}(p)=-S^{f}_n(-p)$.
We utilized these wave functions in Sec.~\ref{sec:Boltzmann}.

At the first order of the expansion the leading correction to the
propagator is a shift of the eigenvalue given by
$\Sigma_{R\,n}(p_\parallel) := \langle\qmn|\Sigma_{R}|\qmn\rangle$.  
Precisely speaking, we need to diagonalize the matrix,
$\langle p_{0},p_{z},\ph, n, \qm,s|\Sigma_{R}|p_{0},p_{z},\ph, n, \qm',s'\rangle$
because of degeneracy of Landau levels.  Here, we assume a convention
after diagonalization, so that the matrix has diagonal components
independent of $\qm$ and $s$.  Then, we can write the propagator up to
the first order as
\begin{equation}
S_{R}(p,p') = S_R(p) (2\pi)^4 \delta^{(4)}(p-p') =\sum_{\qmn}
\langle p| \qmn\rangle \frac{1}{\lambda_{\qmn} + \Sigma_{R\,n}(p_\parallel)}
\overline{\langle\qmn|}p'\rangle\,.
\label{eq:dressedprop}
\end{equation}
The same expression holds for the advanced propagator with
$\Sigma_{R\,n}(p_\parallel)$ replaced with $\Sigma_{A\,n}(p_\parallel)$.

\subsection{Resummation of the vertex function and  the linearized Boltzmann equations}

We are ready to calculate the longitudinal conductivity from the Kubo
formula or more specifically Eq.~\eqref{eq:Kubo}.  In the same way as
in Eq.~\eqref{eq:trick}, in the present calculation, we can slightly
simplify Eq.~\eqref{eq:Kubo} as
\begin{equation}
  \sigma_{\parallel} = \lim_{k_{0}\to0}\lim_{\bk\to\bzero}
  \frac{1}{\ri k_{0}}\im \Pi_{R}^{33}(k)\,.
\end{equation}
Now, we need to evaluate $\Pi_R^{\mu\nu}(k)$ which involves pinch
singularities as we explicitly see here.

The polarization tensor~\eqref{eq:Pioneloop} at the one loop level can
be generalized to the following nonperturbative form:
\begin{equation}
  \Pi_{R}^{\mu\nu}(k) =  (-1)\ri q^{2}\int\frac{\rd^{4}p}{(2\pi)^{4}}
  \tr\bigl[ \Gamma_{A\beta\alpha}^{(0)\mu}(-k,-p,p+k)
  S^{\alpha\alpha'}(p+k)\Gamma_{R\alpha'\beta' }^{\nu}(k,-p-k,p) S^{\beta'\beta}(p) \bigr]\,,
\end{equation}
where $\Gamma_{R/A\,\alpha\beta}^{(0)\mu}$ and
$\Gamma_{R/A\,\alpha\beta}^{\mu}$ are the bare and the dressed
retarded/advanced vertex functions.  As discussed in the beginning of
Sec.~\ref{sec:longitudinal}, the product of retarded and advanced
propagators, $S_{R}(p)S_{A}(p)$, causes infrared enhancement due to
pinch singularities.  We pick this combination of $S_R(p)S_A(p)$
from the diagram, so that we can  approximate $\Pi_{R}^{\mu\nu}(k) $ as
\begin{align}
  \Pi_{R}^{\mu\nu}(k)  &\simeq  (-1)\ri q^{2}(-\ri )^{2}
  \int\frac{\rd^{4}p}{(2\pi)^{4}}\tr\bigl[
  \Gamma_{ARR}^{(0)\mu}(-k,-p,p+k) S_{R}(p+k)
  \Gamma_{RAA}^{\nu}(k,-p-k,p) S_{A}(p) \bigr] \notag\\
& =  (\rme^{\beta k_{0}}-1)\ri q^{2} \int\frac{\rd^{4}p}{(2\pi)^{4}}\,
  n_{F}(p_{0}+k_{0})[1-n_{F}(p_{0})] \notag\\
&\qquad\qquad\qquad\qquad\qquad\times
\tr\bigl[ \gamma^{\mu }S_{R}(p+k)\Gamma_{RAA}^{\nu}(k,-p-k,p) S_{A}(p) \bigr]\,.
\end{align}
Here, we substituted,
\begin{equation}
\Gamma_{ARR}^{(0)\mu}(k_{1},k_{2},k_{3})=
 \bigl(\rme^{-\beta k_{10}}-1\bigr)n_{F}(k_{20})n_{F}(k_{30})\gamma^{\mu}\,,
\end{equation}
deduced from Eq.~\eqref{eq:vertexPhoton}.  Now, in this basis, we can
easily arrive at the following form of the expression for the
longitudinal conductivity:
\begin{equation}
  \sigma_{\parallel}
  \simeq \beta q^{2}\,\re \int\frac{\rd^{4}p}{(2\pi)^{4}}
  n_{F}(p_{0})[1-n_{F}(p_{0})] \tr\bigl[
   \gamma^{3 }S_{R}(p)\Gamma^{3}(p)S_{A}(p)\bigr]\,,
\end{equation}
where we introduced
$\Gamma^{\mu}(p):=\Gamma_{RAA}^{\nu}(0,-p,p)$ to simplify the
notation.

We need to evaluate $S_{R}(p)\Gamma^{3}(p)S_{A}(p)$, in which both propagator and the vertex functions are dressed.
We shall first consider the case with $\Gamma^{3}(p)=\gamma^{3}$ to see the behavior of the pinch singularity.
Using Eq.~\eqref{eq:dressedprop}, we can write down an explicit
integral of $S_{R}(p) \gamma^{3}S_{A}(p)$ as
\begin{align}
  &S_{R}(p) \gamma^{3}S_{A}(p)
  =\int\frac{\rd^{4}p'}{(2\pi)^{4}}\int\frac{\rd^{4}p''}{(2\pi)^{4}}\,
    (2\pi)^{4}\delta^{(4)}(p-p')(2\pi)^{4}\delta^{(4)}(p'-p'')\,
    S_{R}(p) \gamma^{3}S_{A}(p') \notag\\
  &= \sum_{\qmn,\qmn'} \int\frac{\rd^{4}p'}{(2\pi)^{4}}\int\frac{\rd^{4}p''}{(2\pi)^{4}}
   \langle p|  \qmn\rangle
   \frac{1}{\lambda_{\qmn} \!+\!\Sigma_{R\,n}(p_{\parallel})}
   \overline{\langle\qmn|} p' \rangle 
   \gamma^{3} \langle p' |\qmn'\rangle
   \frac{1}{\lambda_{\qmn'} \!+\! \Sigma_{A\,n'}(p_{\parallel})}
   \overline{ \langle \qmn'|} p'' \rangle\,.
\end{align}
With the wave functions~\eqref{eq:pn} we can evaluate the matrix
element as
\begin{equation}
  \int \frac{\rd^{4}p'}{(2\pi)^{4}}\,
  \overline{\langle {\qmn}|} p'\rangle \gamma^{3}\langle p' |\qmn'\rangle  =
  \overline{\langle {\qmn}|} \gamma^{3} |\qmn'\rangle
  =  \frac{p_{z}}{\varepsilon_{n}}\delta_{\qmn,\qmn'} + \cdots,
\end{equation}
where the ellipsis, $\cdots$, denotes the off-diagonal components with
$\qmn\neq \qmn'$.  We note that $p_z$ is a momentum contained in
$\qmn$, but with $\langle p|\qmn\rangle$, we can identify $p_z$ as the
same one in the argument of $S_{R/A}(p)$ [see Eq.~\eqref{eq:pn}].
These off-diagonal components do not contribute to the pinch
singularities, and thus, we can safely neglect them.  We eventually
simplify $ S_{R}(p) \gamma^{3 }S_{A}(p)$ dropping the off-diagonal
components as
\begin{equation}
  S_{R}(p) \gamma^{3 }S_{A}(p) 
  \simeq \sum_{\qmn}
    \int\frac{\rd^{4}p''}{(2\pi)^{4}}\, \frac{p_{z}}{\varepsilon_{n}}\,
    \langle p|\qmn\rangle 
    \frac{1}{\lambda_{\qmn}+\Sigma_{R\,n}(p_{\parallel})} \cdot
    \frac{1}{\lambda_{\qmn}+\Sigma_{A\,n}(p_{\parallel})}
    \overline{\langle \qmn|} p''\rangle\,.
\end{equation}
We note that from the spectral function~\eqref{eq:spectral} the
following relation holds,
\begin{equation}
  \begin{split}
    &\frac{1}{-2\im \Sigma_{R\,n}(p_\parallel)}
    S_n(p) (2\pi)\sgn(p_0)\delta(p_0^2-\varepsilon_n^2)\,
  (2\pi)^4\delta(p-p'') \\
  &\qquad\qquad\qquad\qquad
  \simeq \sum_{\qmn}{}' \langle p|\qmn\rangle \frac{1}
  {\bigl[\lambda_{\qmn}+\Sigma_{R\, n}\bigr]
    \bigl[\lambda_{\qmn}+\Sigma_{A\,n}\bigr]} \overline{\langle\qmn|}p''\rangle\,,
 \end{split}
\end{equation}
apart from the energy shift by $\Sigma_{R/A\,n}$ of higher order.
Here, the prime denotes the sum over $\qmn$ except $n$.
Substituting this into the above, we can find the following form:
\begin{equation}
  S_{R}(p) \gamma^{3 }S_{A}(p) 
 = \sum_{n} \frac{p_{z}}{\varepsilon_n}\cdot\frac{1}{-2 \im \Sigma_{Rn}(p_{\parallel})}
S_{n}(p) (2\pi)\sgn(p_{0})\delta(p_{0}^{2}-\varepsilon_{n}^{2}).
\label{eq:SAGammaSR}
\end{equation}
It is clear from this form that the weak coupling limit,
$\im \Sigma_{R\,n}(p_{\parallel})\to0$, makes
$S_{R}(p) \gamma^{3}S_{A}(p)$ diverge, which is nothing but the pinch
singularity.  We also note that
$\overline{\langle {\qmn}|} \gamma^{1,2} |\qmn'\rangle$ do not have
the diagonal component, so that no pinch singularity appears in the
transverse and Hall conductivities, which justifies our calculations in
Sec.~\ref{sec:transverse}.

The above exercise of $S_R(p)\gamma^3 S_A(p)$ implies the following
reasonable Ansatz for the Dirac structure of the vertex function as
\begin{equation}
S_{R}(p)\, q\Gamma^3(p \,)S_{A}(p) =
\sum_{n} \tilde{\chi}(p_{\parallel})
S_{n}(p) (2\pi)\sgn(p_{0})\delta(p^{2}_{0}-\varepsilon_{n}^{2})\,.
\label{eq:srgsa}
\end{equation}
We note that $\tilde{\chi}(p_\parallel)$ is also an $n$ dependent
function in an implicit way through $p_0$.  With this form we can
treat the transverse momentum integral with $\tr[\gamma^3 S_n(p)]$ in
the same way as in Eq.~\eqref{eq:trans_integ}, and then the
longitudinal conductivity looks,
\begin{equation}
\begin{split}
\sigma_{\parallel}
&= \beta \Nc \frac{q^{2}B}{2\pi} \sum_{n}\alpha_{n}
\int\frac{\rd^{2}p_{\parallel}}{(2\pi)^{2}} \,2p_z\, 
n_{F}(p_0)\bigl[1-n_{F}(p_0)\bigr] (2\pi)\sgn(p_{0})
\delta(p_{0}^{2}-\varepsilon_{n}^{2})\,\tilde{\chi}(p_\parallel) \,.
\end{split}
\end{equation}
We can finally perform the $p_{0}$ integration to find,
\begin{equation}
  \sigma_{\parallel} = \beta \Nc \frac{q^{2}B}{2\pi} \sum_{n}\alpha_{n}
  \int\frac{\rd p_{z}}{2\pi}\, \frac{p_{z}}{\varepsilon_{n}}\,
  n_{F}(\varepsilon_n)\bigl[1-n_{F}(\varepsilon_n)\bigr]
 (\chi_{p}-\bar{\chi}_{p}) \,,
\end{equation}
where we introduced a notation,
$\chi_{p}=\tilde{\chi}(\varepsilon_{n},p_{z})$ and
$\bar{\chi_{p}}=-\tilde{\chi}(-\varepsilon_{n},-p_{z})$.
This final expression exactly corresponds to the second line of
Eq.~\eqref{eq:cond}.

The remaining task is to derive a condition to fix $\chi_p$ and
$\bar{\chi}_p$, which needs the determination of the vertex function.
At the leading order the Bethe-Salpeter equation shown in
Fig.~\ref{fig:BSequation} reads:
\begin{equation}
\begin{split}
  \Gamma^\mu(p)
  &= \gamma^\mu+  (\ri)^{2} \int \frac{\rd^{4}p'}{(2\pi)^{4}}\,
  g_{\gamma A\alpha }(k,-p,p') \,t^{a}\, g_{\gamma' \beta A}(-k,-p',p)
  \,t^{a}\, G^{\gamma\gamma'}_{\rho\nu}(k) \\
  &\qquad\qquad\qquad\qquad\times\gamma^{\rho}S^{\alpha\alpha'}(p')
  \Gamma_{R\alpha'\beta'}^\mu(0,-p',p')S^{\beta'\beta}(p')\gamma^{\nu} \\
  &\simeq \gamma^\mu + C_{F} \int \frac{\rd^{4}p'}{(2\pi)^{4}}\,
  g_{\gamma AR }(k,-p,p')g_{\gamma' R A}(-k,-p',p)
  G^{\gamma\gamma'}_{\rho\nu}(k) \gamma^{\rho}S_{R}(p')
  \Gamma^\mu(p')S_{A}(p')\gamma^{\nu} ,
\end{split}
\end{equation}
where $k=p-p'$ and $C_{F}=(\Nc^{2}-1)/(2\Nc)$.  From the first to the
second lines the approximation symbol, $\simeq$, indicates a
prescription to pick up the pinch singularity only.  From the Feynman
rule~\eqref{eq:vertexGluon}, the product of the coupling can be
evaluated as
\begin{equation}
\begin{split}
  g_{RAR}(k,-p,p')  g_{ARA}(-k,-p',p)
  &=  g^{2} \Bigl\{
  \bigl[1+n_{B}(k_{0})\bigr]\bigl[1-n_{F}(p'_{0})\bigr]+n_{B}(k_{0})n_{F}(p'_{0})
 \Bigr\}\,,\\
  g_{RRA}(k,-p,p') g_{AAR}(-k,-p',p)
  &= -g^{2} \Bigl\{
  \bigl[1+n_{B}(k_{0})\bigr]\bigl[1-n_{F}(p'_{0})\bigr]+n_{B}(k_{0})n_{F}(p'_{0})
 \Bigr\}\,.
\end{split}
\end{equation}
Then, these expressions reduce the Bethe-Salpeter equation into a
simple form as
\begin{equation}
\begin{split}
  \gamma^\mu
  &= \Gamma^\mu(p) - \frac{g^{2}C_{F}}{n_{F}(p_{0})\bigl[1-n_{F}(p_{0})\bigr]}
  \int\frac{\rd^{4}p'}{(2\pi)^{4}} \bigl[1+n_{B}(k_{0})\bigr]
  \bigl[1-n_{F}(p'_{0})\bigr] n_{F}(p_{0}) \rho_{\rho\nu}(k)\\
  &\qquad\qquad \qquad\qquad \qquad\qquad \qquad\qquad
  \times\gamma^{\rho}S_{R}(p') \Gamma^\mu(p')S_{A}(p')\gamma^{\nu}\,.
\label{eq:BSEq}
\end{split}
\end{equation}
Here, in the above, we interchanged $\Gamma^{\mu}$ and $\gamma^{\mu}$
from the left- and the right-hand sides for convenience and flipped
the sign.  The gluon propagator amounts to the spectral function,
$\rho_{\rho\nu}(k):=G^{RA}_{\nu\rho}(k)-G^{AR}_{\nu\rho}(k)
=-\ri [G_{R\nu\rho}(k)-G_{A\nu\rho}(k)]$.  At the leading order of
coupling constant, in the Feynman gauge, the spectral function reads,
\begin{equation}
\rho_{\rho\nu}(k)=-\eta_{\rho\nu}(2\pi)\sgn(k_{0})\delta(k^{2}).
\label{eq:spect_Feyn}
\end{equation}
Then, we aim to translate the condition~\eqref{eq:BSEq} for
$\chi_p$ and $\bar{\chi}_p$ contained in $\Gamma^\mu(p')$ into the
form of linearized Boltzmann equations.  The explicit relation between
the vertex function and $\tilde{\chi}$ is given by
Eq.~\eqref{eq:srgsa}, from which we can easily solve $\Gamma^3(p')$ as
\begin{align}
q\Gamma^{3}(p') &= 
S_{R}(p')^{-1}\sum_{n} \tilde{\chi}(p_{\parallel}') S_{n'}(p')
(2\pi)\sgn(p'_{0})\delta(p'^{2}_{0}-\varepsilon_{n'}^{2}) S_{A}(p')^{-1} \notag\\
&=  -\ri\bigl[ S_{A}(p')^{-1}- S_{R}(p')^{-1}\bigr]\,\tilde{\chi}(p_{\parallel}) \notag\\
&=  -2\im\Sigma(p)\,\tilde{\chi}(p_{\parallel}) \,.
\label{eq:vertexChi}
\end{align}
Multiplying the Bethe-Salpeter equation~\eqref{eq:BSEq} by
$qS_{n}(p)$, taking the trace, and using Eqs.~\eqref{eq:srgsa}, 
\eqref{eq:spect_Feyn},  and \eqref{eq:vertexChi}, we find,
\begin{equation}
  \begin{split}
  &q\Nc\,\tr \bigl[ S_{n}(p) \gamma^3 \bigr]
    =  -  \Nc \tr\bigl[ S_{n}(p)2 \im \Sigma_{R}(p))\bigr]\tilde{\chi}(p_\parallel) \\
  &\qquad\quad
    - \frac{g^{2}C_{F}\Nc}{n_{F}(p_{0})\bigl[1-n_{F}(p_{0})\bigr]}
  \sum_{n'}\int \frac{\rd^{4}p'}{(2\pi)^{4}} \bigl[1+n_{B}(k_{0})\bigr]
  \bigl[1-n_{F}(p'_{0})\bigr] n_{F}(p_{0}) \\
  &\qquad\quad
    \times(2\pi)\sgn(k_{0})\delta(k^{2})(2\pi)\sgn(p'_{0})\delta(p'^{2}_{0}-\varepsilon_{n'}^{2})
    \bigl\{- \tr\bigl[ S_{n}(p)\gamma_{\nu}S_{n'}(p') \gamma^{\nu} \bigr]
    \tilde{\chi}(p'_{\parallel}) \bigr\}\,.
  \end{split}
  \label{eq:BSvertex}
\end{equation}
The direct evaluation of the self-energy yields,
\begin{align}
2\,\im \,\Sigma_{R}(p) &= 2\,\im\,(\ri)^{3}\!\int\frac{\rd^{4}p'}{(2\pi)^{4}}\,
G_{{\alpha'}\alpha;\mu\nu}(k)
g_{\alpha' A\beta'}(k,-p,p')\gamma^{\mu} t^{a}
S_{\beta'\beta}(p') g_{\alpha \beta R}(-k,-p',p)\gamma^{\nu}t^{a} \notag\\
&=2\,\re \,C_{F}\!\int\frac{\rd^{4}p'}{(2\pi)^{4}}\Bigl[
G_{{A}\mu\nu}(k)
g_{A AR}(k,-p,p')\gamma^{\mu} 
S_{R}(p') g_{R A R}(-k,-p',p)\gamma^{\nu} \notag\\
&\qquad\qquad\qquad\qquad+G_{{R}\mu\nu}(k)
g_{R AA}(k,-p,p')\gamma^{\mu} 
S_{A}(p') g_{A R R}(-k,-p',p)\gamma^{\nu} \notag\\
&\qquad\qquad\qquad\qquad+G_{{R}\mu\nu}(k)
g_{R AR}(k,-p,p')\gamma^{\mu} 
S_{R}(p') g_{A A R}(-k,-p',p)\gamma^{\nu} \Bigr] \notag\\
&=\frac{g^{2}C_{F}}{n_{F}(p_{0})\bigl[1-n_{F}(p_{0})
\bigr]}\sum_{n'}\int\frac{\rd^{4}p'}{(2\pi)^{4}}
\bigl[1+n_{B}(k_{0})\bigr]\bigl[1-n_{F}(p'_{0})\bigr]n_{F}(p_{0}) \notag\\
&\qquad\times  (2\pi)\sgn(k_{0})\delta(k^{2})  (2\pi)\sgn(p_{0}')
\delta(p'^{2}_{0}-\varepsilon_{n'}^{2}) \gamma_{\nu}  S_{n'}(p')\gamma^{\nu}\,.
\end{align}
Interestingly, this form has similarity to the second term in
Eq.~\eqref{eq:BSvertex}, so that we can further simplify the
Bethe-Salpeter equation into the following form:
\begin{align}
& \Nc\frac{p_{z}}{p_{0}}\tr \bigl[\gamma^{0}S_{n}(p) \bigr] = -
 \frac{g^{2}C_{F}\Nc}{n_{F}(p_{0})\bigl[1-n_{F}(p_{0})\bigr]}
    \sum_{n'}\int \frac{\rd^{4}k}{(2\pi)^{4}}\int \frac{\rd^{4}p'}{(2\pi)^{4}}
    (2\pi)^{4}\delta^{(4)}(p-k-p') \notag\\
  & \times \tr\bigl[ S_{n}(p) \gamma_{\nu}S_{n'}(p')\gamma^{\nu} \bigr]
  \, n_{F}(p_{0})\bigl[1\!-\!n_{F}(p'_{0})\bigr]
    \bigl[1\!+\!n_{B}(k_{0})\bigr]
    (2\pi)\sgn(k_{0})\delta(k^{2}) \notag\\
  &\times (2\pi)\sgn(p'_{0})\delta(p'^{2}_{0}\!-\!\varepsilon_{n'}^{2})
 \bigl[\tilde{\chi}(p_\parallel)-\tilde{\chi}(p'_{\parallel}) \bigr],
 \label{eq:LinearizedBoltzmannDiagrammatic}
\end{align}
where we employed $ \tr[\gamma^{3}S_{n}(p)]=
(p_{z}/p_{0})\tr[\gamma^{0}S_{n}(p)]$ to rewrite the left-hand side.

To confirm that the above expression is nothing but the Boltzmann
equation, we shall focus on the positive energy state, i.e.,
$p_{0}=\varepsilon_{n}>0$.  We perform the $k_0$ and $p_0'$
integrations in the left-hand side of
Eq.~\eqref{eq:LinearizedBoltzmannDiagrammatic}, and then, the
expression takes the following form,
\begin{equation}
\begin{split}
&\Nc\frac{p_{z}}{\varepsilon_{n}}\tr \bigl[\gamma^{0}S_{n}(p)\bigr]\\
&=\frac{1}{n_{F}(\varepsilon_{n})\bigl[1-n_{F}(\varepsilon_{n})\bigr]} \sum_{n'}\int \frac{\rd^{3}k}{(2\pi)^{3}}\frac{1}{2|\bk|}\int \frac{\rd^{3}p'}{(2\pi)^{3}}\frac{1}{2\varepsilon_{n'}}\\
&\quad\times\biggl\{
\overline{|\mathcal{M}_{p\to p'+k}|}(2\pi)^{4}\delta^{(4)}(p-k-p') n_{F}(\varepsilon_{n})[1-n_{F}(\varepsilon_{n'})][1+n_{B}(|\bk|)]
(\chi_{p}-\chi_{p'} )\\
&\qquad+\overline{|\mathcal{M}_{p+k\to p'}|}(2\pi)^{4}\delta^{(4)}(p+k-p') n_{F}(\varepsilon_{n})n_{B}(|\bk|)[1-n_{F}(\varepsilon_{n'})]
(\chi_{p}-\chi_{p'}) \\
&\qquad+\overline{|\mathcal{M}_{p+p'\to k}|}(2\pi)^{4}\delta^{(4)}(p-k+p') n_{F}(\varepsilon_{n})n_{F}(\varepsilon_{n'})[1+n_{B}(|\bk|)]
(\chi_{p}+\bar{\chi}_{p'} )
\biggr\}. \label{eq:LinearizedBoltzmannEquation}
\end{split}
\end{equation}
Here, we defined the squared matrix element averaging over all
internal degrees of freedom as
\begin{equation}
\begin{split}
  \overline{|\mathcal{M}_{p\to p'+k}|} &:=
  -g^{2}C_{F}\Nc\, \tr\bigl[ S_{n}(p) \gamma_{\nu}S_{n'}(p')\gamma^{\nu}\bigr]
  \Bigr|_{p_{0}'=\varepsilon_{n'},k_{0}=|\bk|} \;,\\
  \overline{|\mathcal{M}_{p+k\to p'}|} &:=
  -g^{2}C_{F}\Nc\, \tr\bigl[ S_{n}(p) \gamma_{\nu}S_{n'}(p')\gamma^{\nu}\bigr]
  \Bigr|_{p_{0}'=\varepsilon_{n'},k_{0}=|\bk|} \;,\\
  \overline{|\mathcal{M}_{p+p'\to k}|} &:=
  g^{2}C_{F}\Nc\, \tr\bigl[ S_{n}(p) \gamma_{\nu}S_{n'}(-p')\gamma^{\nu}\bigr]
  \Bigr|_{p_{0}'=\varepsilon_{n'},k_{0}=|\bk|} \;.
\end{split}
\end{equation}
Now, we see that Eq.~\eqref{eq:LinearizedBoltzmannEquation} is nothing
but the linearized Boltzmann equation as considered in
Sec.~\ref{sec:longitudinal}.  This completes our discussions on the
diagrammatic derivation of the Boltzmann equations.


\section{Conclusions}
\label{sec:summary}

We spelled out all the details about the field-theoretical calculation
of the electric conductivity beyond the lowest Landau level
approximation.  Including higher Landau level contributions the
electric conductivity turns out to behave smoothly as a function of
fermion mass, and particularly, we found that the electric
conductivity remains finite even in the massless limit.  The diverging
part of the electric conductivity from the lowest Landau level is
attributed to the chiral magnetic effect which predicts a finite
production rate of topological current, that is, a topological current
proportional to time.  This current is scattered off with higher
Landau levels at finite mass, so that a steady state is balanced with
current not blowing up, and we can obtain a physically sensible
estimate for the electric conductivity.

Contrary to the common statement that the chiral magnetic effect leads
to quadratic suppression of the negative magnetoresistance, our
results (in which the axial charge in the massless limit is not
treated as a hydrodynamic mode, which is needed for the CME) show
that the asymptotic behavior of the electric conductivity is linearly
suppressed with increasing magnetic field.  In fact, we found that in
an intermediate region of the applied magnetic field, there is a
window where the magnetic dependence of the electric conductivity
looks quadratic approximately.  This crucial difference from the
previous argument based on the relaxation time approximation appears
from potential magnetic dependence of the relaxation time.

Technically speaking, our formulation itself contains some notable
features even apart from complications associated with the Landau
level sum.  We made resummation to deal with the pinch
singularities, for which we started with the Bethe-Salpeter
equation.  The most cumbersome part in the calculations lies in the
field-theoretical reduction from the Bethe-Salpeter equation into the
Boltzmann equation.  In the present paper, we first presented the
electric conductivity results presuming the Boltzmann equation, and
later on, we gave full calculations based on the Schwinger-Keldysh
formalism to justify our kinetic treatments.  At some point of our
calculations one might have wondered a relation to what has been
discussed in the context of the chiral kinetic theory.  We note here
that, when we refer to the Boltzmann equation in this work, we retain
the Dirac matrix structures.  In the chiral kinetic theory, in
contrast, the Dirac structure or the spin dynamics is encoded in the
Berry curvature in the adiabatic approximation.  In this sense we can
say that nothing is dropped off from our present formulation keeping
the Dirac matrices, and the effect of the chiral anomaly must be fully
incorporated even though the Berry curvature does not manifest itself
here.

We assumed fundamental dynamics of quarks and gluons in this work, but
it should be feasible to apply our methodology in order to explain the table-top
experiment of the negative magnetoresistance.  For this purpose,
unlike QCD that is clearly defined by a simple Lagrangian, we need to
adopt some impurity model and introduce interactions between gapless
fermions and phonons.  It would be then a challenging problem to
attempt theoretical calculations in order to understand quantitative
features of the negative magnetoresistance such as the detailed shape
of magnetic dependence, the temperature dependence of the
magnetoresistance, similarity and difference depending on various
chiral materials, and so on.  More and more exciting developments in
theory will be awaiting us, and we believe that the present work takes
the initiative in those directions of future research.

\acknowledgments

The authors thank, Dima~Kharzeev,
Koichi~Hattori, and
Daisuke~Satow
for discussions.
This work was supported in part by the ExtreMe Matter Institute EMMI
at the GSI Helmholtzzentrum f\"{u}r Schwerionenphysik, Darmstadt, Germany.
K.~F.\ is grateful for warm hospitality at Heidelberg University where
a part of this work was complete. 
This work was also supported by Japan Society for the Promotion of Science
(JSPS) KAKENHI Grant
No.\ 
16K17716, 17H06462, and 18H01211.

\bibliographystyle{JHEP}
\bibliography{magnetic}

\end{document}